# Coupled Cluster-Inspired Geminal Wavefunctions

Pratiksha B. Gaikwad,[1] Taewon D. Kim,[1] M. Richer,[2] Rugwed A. Lokhande,[1] Gabriela Sánchez-Díaz,[2] Peter A. Limacher,[2] Paul W. Ayers,[2*] Ramón Alain Miranda-Quintana[1*]

1. Department of Chemistry and Quantum Theory Project, University of Florida, Gainesville, FL 32603, USA
2. Department of Chemistry and Chemical Biology, McMaster University, Hamilton, Ontario, L8S 4M1, Canada

Emails: quintana@chem.ufl.edu, ayers@mcmaster.ca

**Abstract**

Electron pairs have an illustrious history in chemistry, from powerful concepts to understanding structural stability and reactive changes, to the promise of serving as building blocks of quantitative descriptions of the electronic structure of complex molecules and materials. However, traditionally, two-electron wavefunctions (*geminals*), have not enjoyed the popularity and widespread use of the more standard single-particle methods. This has changed recently, with a renewed interest in the development of geminal wavefunctions as an alternative to describing strongly-correlated phenomena. Hence, there is a need to find geminal methods that are accurate, computationally-tractable, and do not demand significant input from the user (particularly, via cumbersome and often ill-behaved orbital optimization steps). Here we propose new families of geminal wavefunctions, inspired by the pair Coupled Cluster (pCCD) ansatz. We present a new hierarchy of two-electron wavefunctions that extend the one-reference orbital idea to other geminals. Moreover, we show how to incorporate single-like excitations in this framework, without leaving the quasiparticle picture. We explore the role of imposing seniority restrictions on these wavefunctions, and benchmark these new methods on model strongly-correlated systems.

## 1. INTRODUCTION

Single-particle states (e.g., orbitals) have been pivotal in the development of modern quantum and computational chemistry.[1,2] From tools to interpreting chemical reactions and bonding,[3-8] to building blocks of density[9,10] and reduced-density-based[11-19] methods, the simple picture of electrons occupying definite states is a very powerful abstraction. This notion naturally leads to the concept of electron configuration (e.g., a Slater determinant), which pervades all *ab initio* wavefunction development. Each such determinant describes electrons in (quasi-)independent states, with the lowest energy configuration corresponding to the Hartree-Fock (HF) wavefunction,[20] $|\Phi_0\rangle$. This mean-field approach can then be expanded to include the missing contributions from the electron-electron interactions: the *electron correlation*.[21] In cases where $|\Phi_0\rangle$ provides a good enough approximation to the actual wavefunction (e.g., *weakly correlated* systems), standard methods like Configuration Interaction (CI),[1,2,22-25] Coupled Cluster (CC),[26-28] or Perturbation Theory (PT)[27,29-32] can capture a large fraction of the missing electron correlation. However, in some situations, the electron-electron interaction energy is larger than the energy gaps between the orbitals (a behavior that is exacerbated when there is a large number of quasi-degenerate states).[33] It is customary to refer to these systems as strongly correlated, and in these cases, we cannot typically identify a dominant configuration,[34] so most single-reference methods fail to provide even qualitatively correct results. (Albeit, in some cases, this can be alleviated, as illustrated below.)[35]

Dealing with strongly-correlated systems is one of the biggest challenges in modern theoretical chemistry. Here, we will be concerned with one particularly promising line of attack to this problem: using two-electron states (*geminals*)[36-55] as the main building blocks of the wavefunction. This framework can be easily presented if we establish some parallels with the orbital picture. First, we need to define a geminal creation operator, $G_p^\dagger$, which can be represented as:

$$G_p^\dagger = \sum_{r,q} c_{p;rq} a_r^\dagger a_q^\dagger \qquad (1)$$

with $a_r^\dagger$ being the standard spin-orbital creation operators. Then, we just build a mean-field wavefunction using these electron pairs:

$$|\Psi_{APG}\rangle = \prod_{p=1}^{N/2} G_p^\dagger |\theta\rangle = \prod_{p=1}^{N/2} \left( \sum_{r,q} c_{p;rq} a_r^\dagger a_q^\dagger \right) |\theta\rangle \qquad (2)$$

where $|\theta\rangle$ is the physical vacuum, and we explicitly indicate that we are working with $N$ electrons. Eq. (2) represents the most general geminal wavefunction: the Antisymmetric Product of Geminals (APG),[56] which includes every possible pairing between spin-orbitals and does not impose any restrictions on the geminal coefficients $c_{p;rq}$.

The APG wavefunction is very accurate but, unfortunately, it is too computationally costly.[57] Hence, it is necessary to introduce several approximations while building the geminals, in order to make them tractable. Arguably the simplest choice is to use the same geminal to describe all the electrons in the system:

$$|\Psi_{AGP}\rangle = \left(G^\dagger\right)^{\frac{N}{2}}|\theta\rangle = \left(\sum_r c_r a_r^\dagger a_{\bar{r}}^\dagger\right)^{\frac{N}{2}}|\theta\rangle \tag{3}$$

(Notice that we have indicated the paired partner of spin-orbital $r$ by a bar on top, $\bar{r}$). This Antisymmetrized Geminal Power[54,58-61] (AGP) wavefunction is very limited on its own, but it can be corrected with Jastrow factors[62-64] to improve its accuracy. It has also gained popularity recently as a convenient starting point for describing systems with strong pairing interactions[65-69] in classical and quantum computers.[70,71] Likewise, strongly orthogonal geminals (as in the Antisymmetrized Product of Strongly Orthogonal Geminals, APSG):[43,72-80]

$$|\Psi_{APSG}\rangle = \prod_{p=1}^{N/2} G_p^\dagger |\theta\rangle = \prod_{p=1}^{N/2}\left(\sum_r c_{p;r} a_{p_r}^\dagger a_{\bar{p}_r}^\dagger\right)|\theta\rangle; p \neq q \Rightarrow \{a_{p_r}^\dagger\} \cap \{a_{q_r}^\dagger\} = \varnothing \tag{4}$$

can dramatically reduce the computational burden by building geminals with non-overlapping sets of spin-orbitals and can later on be improved using perturbative techniques.[13,72,81]

While promising, APG and APSG demand multiple corrections in order to properly describe realistic systems. For this reason, there has been a renewed interest in designing interacting geminal wavefunctions that limit the possible pairings between the spin-orbitals, but not as drastically as APG and APSG. For example, in the Antisymmetric Product of Set-separated Geminals (APsetG)[82] the spin-orbitals are separated in two disjoint sets, and no pairs with both elements from the same set can be formed:

$$|\Psi_{APsetG}\rangle = \prod_{p=1}^{N/2} G_p^\dagger |\theta\rangle = \prod_{p=1}^{N/2}\left(\sum_{\substack{r\in S_1, q\in S_2 \\ S_1\cap S_2=\varnothing}} c_{p;rq} a_r^\dagger a_q^\dagger\right)|\theta\rangle \tag{5}$$

We can impose even more restrictions, as in the Antisymmetric Product of Interacting Geminals (APIG), [51,52,83] which only allows for an alpha/beta spin-orbital to be paired with its corresponding beta/alpha companion:

$$|\Psi_{APIG}\rangle = \prod_{p=1}^{N/2} G_p^\dagger |\theta\rangle = \prod_{p=1}^{N/2} \left( \sum_r c_{p;r} a_r^\dagger a_{\bar{r}}^\dagger \right) |\theta\rangle \qquad (6)$$

The graphs shown in Fig. 1 illustrate the simple hierarchical relation between APG, APsetG, and APIG, with the former being the most complete wavefunction, while the latter is clearly the simplest of these three methods. The problem, however, is that even APIG is intractable. The easiest way to see this is by realizing that the overlap of the APIG wavefunction with a Slater determinant is expressed as a permanent of the geminal coefficients, and computing this function is a #P-hard problem.[84]

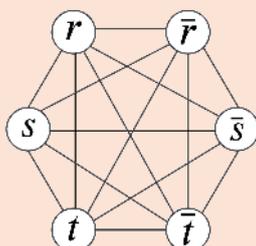

**Figure 1.** Geminal creation operator $G_p^\dagger$ and spin orbital pairing schemes in geminal for APG, APsetG and APIG wavefunctions.

There have been two main routes to simplify APIG, aiming to turn it into a practical method. The first one draws inspiration from exactly solvable models[85-89] and imposes conditions on the geminal coefficients. For instance, if one parametrizes the $c_{p;r}$ in Eq. (6) as:

$$c_{p;r} = \frac{\varsigma_r}{\varepsilon_r - \lambda_p} \qquad (7)$$

the resulting permanents can be easily calculated as functions of auxiliary determinants thanks to Borchardt's theorem.[90-93] This choice corresponds to the Antisymmetric Product of rank-2 Geminals (APr2G).[94] There has been considerable progress on using model

Hamiltonians to guide the development of new geminal wavefunctions, but parametrizations like Eq. (7) can be too numerically unstable. On the other hand, one could drastically reduce the number of elements in the geminal coefficient matrix so the calculation of the permanent is greatly simplified. This can be done by introducing an occupied-virtual separation in the geminal picture. For example:

$$|\Psi_{AP1roG}\rangle = \prod_{i=1}^{N/2} G_i^\dagger |\theta\rangle = \prod_{i=1}^{N/2} \left( a_i^\dagger a_{\bar{i}}^\dagger + \sum_a c_{i\bar{i};a\bar{a}} a_a^\dagger a_{\bar{a}}^\dagger \right) |\theta\rangle \quad (8)$$

This is the Antisymmetric Product of 1-reference Orbital Geminals (AP1roG).[94] Notice that (as it is customary) we have use labels $i, j, k, \ldots$ to indicate occupied orbitals, labels $a, b, c, \ldots$ to represent the virtual manifold, while the previously used $p, q, r, \ldots$ denote arbitrary spin-orbitals. This wavefunction has had great success in recent years, being applied to model strongly correlated systems,[37,95] excited states,[96,9798] heavy elements,[99,100] and in embedding studies.[101] Much of this has to do with the fact that we can reformulate AP1roG as a Coupled Cluster (CC) wavefunction, more precisely, as a simplified form of CCD in which only pair double excitations are allowed: pCCD.[102] This is easy to see since:

$$\begin{aligned} |\Psi_{AP1roG}\rangle &= \prod_{i=1}^{N/2} \left( a_i^\dagger a_{\bar{i}}^\dagger + \sum_a c_{i\bar{i};a\bar{a}} a_a^\dagger a_{\bar{a}}^\dagger \right) |\theta\rangle \\ &= \prod_{i=1}^{N/2} \left( 1 + \sum_a c_{i\bar{i};a\bar{a}} a_a^\dagger a_{\bar{a}}^\dagger a_{\bar{i}} a_i \right) a_i^\dagger a_{\bar{i}}^\dagger |\theta\rangle \\ &= \prod_{i=1}^{N/2} \left( 1 + \sum_a c_{i\bar{i};a\bar{a}} a_a^\dagger a_{\bar{a}}^\dagger a_{\bar{i}} a_i \right) |\Phi_0\rangle \\ &= \exp\left( \sum_{i,a} c_{i\bar{i};a\bar{a}} a_a^\dagger a_{\bar{a}}^\dagger a_{\bar{i}} a_i \right) |\Phi_0\rangle \end{aligned} \quad (9)$$

The fact that AP1roG and pCCD are identical immediately implies that this is a tractable method, and as noted before, it opens the way to using CC techniques with this geminal wavefunction. However, the price to pay for introducing the occupied-virtual separation is that we need to be very careful selecting the orbitals that will form the pairs.[103] This is critical to achieving quantitatively accurate results, and even to making AP1roG a size-consistent wavefunction.[104-106] Unfortunately, the orbital optimization (OO) step is not only the computational bottleneck, but it could also be ill-behaved in some cases, with the optimization getting easily trapped in a sea of local minima.

In this contribution we build on the insights provided by AP1roG and show that the 1-reference orbital idea can be used to turn geminal wavefunctions more general than

APIG into tractable methods. Perhaps even more importantly, we show how to essentially perform the OO in a truly black-box way by using single excitations "masquerading" as double-excitation operators, so we do not have to abandon the geminal picture. We explore several modifications to these new methods, based on different ways of breaking the seniority in the occupied and/or virtual spaces, leading to new CC-like methods.

## 2. THEORY

2.1 1-reference orbital geminal wavefunctions

Our first key result is to notice that we can easily generate 1-reference orbital versions of the APG and APsetG wavefunctions. For instance, in the case of APG:

$$
\begin{aligned}
\left|\Psi_{\text{APG1roD}}\right\rangle &= \prod_{i=1}^{N/2}\left(a_i^\dagger a_{\bar{i}}^\dagger + \sum_{a,b} c_{i\bar{i};ab} a_a^\dagger a_b^\dagger\right)|\theta\rangle \\
&= \prod_{i=1}^{N/2}\left(1 + \sum_{a,b} c_{i\bar{i};ab} a_a^\dagger a_b^\dagger a_{\bar{i}} a_i\right) a_i^\dagger a_{\bar{i}}^\dagger |\theta\rangle \\
&= \prod_{i=1}^{N/2}\left(1 + \sum_{a,b} c_{i\bar{i};ab} a_a^\dagger a_b^\dagger a_{\bar{i}} a_i\right)|\Phi_0\rangle \\
&= \exp\left(\sum_{i,a,b} c_{i\bar{i};ab} a_a^\dagger a_b^\dagger a_{\bar{i}} a_i\right)|\Phi_0\rangle
\end{aligned}
\tag{10}
$$

Similarly, for APsetG:

$$
\begin{aligned}
\left|\Psi_{\text{APset1roGD}}\right\rangle &= \prod_{i=1}^{N/2}\left(a_i^\dagger a_{\bar{i}}^\dagger + \sum_{\substack{a\in S_1, b\in S_2 \\ S_1\cap S_2=\varnothing}} c_{i\bar{i};ab} a_a^\dagger a_b^\dagger\right)|\theta\rangle \\
&= \prod_{i=1}^{N/2}\left(1 + \sum_{\substack{a\in S_1, b\in S_2 \\ S_1\cap S_2=\varnothing}} c_{i\bar{i};ab} a_a^\dagger a_b^\dagger a_{\bar{i}} a_i\right) a_i^\dagger a_{\bar{i}}^\dagger |\theta\rangle \\
&= \prod_{i=1}^{N/2}\left(1 + \sum_{\substack{a\in S_1, b\in S_2 \\ S_1\cap S_2=\varnothing}} c_{i\bar{i};ab} a_a^\dagger a_b^\dagger a_{\bar{i}} a_i\right)|\Phi_0\rangle \\
&= \exp\left(\sum_{\substack{i,a\in S_1, b\in S_2 \\ S_1\cap S_2=\varnothing}} c_{i\bar{i};ab} a_a^\dagger a_b^\dagger a_{\bar{i}} a_i\right)|\Phi_0\rangle
\end{aligned}
\tag{11}
$$

As was the case for APIG, by introducing the occupied-virtual separation we have now obtained APG- and APsetG-like geminal flavors that are simplified versions of CCD. Therefore, these new geminal functions are obviously tractable, yet they are more general than pCCD in that they can deal with non-seniority zero states. Notice that since APG is the most general pairing wavefunction possible, and yet it is amenable to a CC representation, this implies that other, simpler, geminal wavefunctions could have CC 1-reference analogues. This strategy is motivated by the success of other single-reference CC methods in tackling strongly-correlated systems. Notable examples of this are the CC(P;Q) approach of Piecuch *et al.*,[107-109] externally corrected CC[110-112] and, most notably, Scuseria's work on singlet-paired CCD[35] (CCD0).

Finally, a comment on notation. In the rest of this manuscript, we will favor the more explicit (and hence, slightly more cumbersome) "geminal naming convention" for the new wavefunctions. However, we will also propose alternative CC names for these methods. For instance, the 1-reference orbital variant of APG is the most complete approximation to CCD that is also a geminal wavefunction, so we propose to call it geminal CCD, or gCCD. In the same vein, the 1-reference orbital APsetG could be called set CCD, or sCCD.

2.2 Introducing single-like excitations in the geminal framework

The APG1roD and APset1roGD wavefunctions also require an OO step, but this is something that we would like to avoid, given the difficulties converging to the appropriate solution. This problem is easy to handle in a traditional CC context, because the single excitations allow the orbitals to relax, since the Thouless Theorem indicates that the $T_1$ operator can modify the reference to any other determinant.[113] However, standard single excitations are 1-body terms so, at least on the surface, they seem to be incompatible with the geminal picture.

As a way around this, we propose to mimic the effect of the $T_1$ operators with $T_2$ excitations. For example, we can augment the geminal operators used in AP1roG, Eq. (8), according to:

$$G^\dagger_{i(\text{AP1roGSD})} = a^\dagger_i a^\dagger_{\bar{i}} + \sum_a c_{i\bar{i};a\bar{a}} a^\dagger_a a^\dagger_{\bar{a}} + \sum_a c_{i\bar{i};a\bar{i}} a^\dagger_a a^\dagger_{\bar{i}} \qquad (12)$$

The key change here is that now we allow for the 1-reference spin-orbital to be paired with virtual spin-orbitals. Then, when we take the product of these new geminals, we can write:

$$|\Psi_{AP1roGSD}\rangle = \prod_{i=1}^{N/2}\left(a_i^\dagger a_{\bar{i}}^\dagger + \sum_a c_{i\bar{i};a\bar{a}} a_a^\dagger a_{\bar{a}}^\dagger + \sum_a c_{i\bar{i};a\bar{i}} a_a^\dagger a_{\bar{i}}^\dagger\right)|\theta\rangle$$

$$= \prod_{i=1}^{N/2}\left(1 + \sum_a c_{i\bar{i};a\bar{a}} a_a^\dagger a_{\bar{a}}^\dagger a_{\bar{i}} a_i + \sum_a c_{i\bar{i};a\bar{i}} a_a^\dagger a_{\bar{i}}^\dagger a_{\bar{i}} a_i\right) a_i^\dagger a_{\bar{i}}^\dagger |\theta\rangle$$

$$= \prod_{i=1}^{N/2}\left(1 + \sum_a c_{i\bar{i};a\bar{a}} a_a^\dagger a_{\bar{a}}^\dagger a_{\bar{i}} a_i + \sum_a c_{i\bar{i};a\bar{i}} a_a^\dagger a_{\bar{i}}^\dagger a_{\bar{i}} a_i\right)|\Phi_0\rangle \qquad (13)$$

$$= \exp\left(\sum_{i,a} c_{i\bar{i};a\bar{a}} a_a^\dagger a_{\bar{a}}^\dagger a_{\bar{i}} a_i + \sum_{i,a} c_{i\bar{i};a\bar{i}} a_a^\dagger a_{\bar{i}}^\dagger a_{\bar{i}} a_i\right)|\Phi_0\rangle$$

We can write the wavefunction in the exponential CC form because:

$$\left(\hat{t}_{i\bar{i}}^{a\bar{a}}\right)^2 = \left(\hat{t}_{i\bar{i}}^{a\bar{i}}\right)^2 = 0$$
$$\hat{t}_{i\bar{i}}^{a\bar{a}} \hat{t}_{i\bar{i}}^{b\bar{b}} = \hat{t}_{i\bar{i}}^{a\bar{a}} \hat{t}_{i\bar{i}}^{b\bar{i}} = \hat{t}_{i\bar{i}}^{a\bar{i}} \hat{t}_{i\bar{i}}^{b\bar{i}} = 0 \qquad (14)$$

with the other operators trivially commuting (for simplicity, we used the notation $\hat{t}_{pq}^{rs} = a_r^\dagger a_s^\dagger a_q a_p$).

Notice that while the new terms, $\hat{t}_{i\bar{i}}^{a\bar{i}} = a_a^\dagger a_{\bar{i}}^\dagger a_{\bar{i}} a_i$ are two-body operators (and thus they fit seamlessly in a geminal theory), they effectively just try to excite a single electron from the occupied spin-orbital *i* to the virtual spin-orbital *a*. There are two ways in which we could allow these "faux-single" excitations to happen:

a) Preserving the $M_z$ quantum number, that is, restricting ourselves to the cases in which *i* and *a* have the same spin. We call this: AP1roGSDspin.
b) Without any restrictions on the spins of *i* and *a*. We call this: AP1roGSDgeneralized.

(In "CC nomenclature", we could call these: pCCSDspin, and pCCSDgeneralized, respectively.)

This same procedure can then be repeated for the APG1roD and APset1roGD wavefunctions:

$$G_{i(APG1roSD)}^\dagger = a_i^\dagger a_{\bar{i}}^\dagger + \sum_{a,b} c_{i\bar{i};ab} a_a^\dagger a_b^\dagger + \sum_a c_{i\bar{i};a\bar{i}} a_a^\dagger a_{\bar{i}}^\dagger \qquad (15)$$

$$|\Psi_{APG1roSD}\rangle = \prod_{i=1}^{N/2}\left(a_i^\dagger a_{\bar{i}}^\dagger + \sum_{a,b} c_{i\bar{i};ab} a_a^\dagger a_b^\dagger + \sum_a c_{i\bar{i};a\bar{i}} a_a^\dagger a_{\bar{i}}^\dagger\right)|\theta\rangle$$

$$= \prod_{i=1}^{N/2}\left(1 + \sum_{a,b} c_{i\bar{i};ab} a_a^\dagger a_b^\dagger a_{\bar{i}} a_i + \sum_a c_{i\bar{i};a\bar{i}} a_a^\dagger a_{\bar{i}}^\dagger a_{\bar{i}} a_i\right) a_i^\dagger a_{\bar{i}}^\dagger |\theta\rangle \quad (16)$$

$$= \prod_{i=1}^{N/2}\left(1 + \sum_{a,b} c_{i\bar{i};ab} a_a^\dagger a_b^\dagger a_{\bar{i}} a_i + \sum_a c_{i\bar{i};a\bar{i}} a_a^\dagger a_{\bar{i}}^\dagger a_{\bar{i}} a_i\right) |\Phi_0\rangle$$

$$= \exp\left(\sum_{i,a,b} c_{i\bar{i};ab} a_a^\dagger a_b^\dagger a_{\bar{i}} a_i + \sum_{i,a} c_{i\bar{i};a\bar{i}} a_a^\dagger a_{\bar{i}}^\dagger a_{\bar{i}} a_i\right)|\Phi_0\rangle$$

$$G^\dagger_{i(APset1roGSD)} = a_i^\dagger a_{\bar{i}}^\dagger + \sum_{\substack{a\in S_1, b\in S_2 \\ S_1\cap S_2=\varnothing}} c_{i\bar{i};ab} a_a^\dagger a_b^\dagger + \sum_a c_{i\bar{i};a\bar{i}} a_a^\dagger a_{\bar{i}}^\dagger \quad (17)$$

$$|\Psi_{APset1roGSD}\rangle = \prod_{i=1}^{N/2}\left(a_i^\dagger a_{\bar{i}}^\dagger + \sum_{\substack{a\in S_1, b\in S_2 \\ S_1\cap S_2=\varnothing}} c_{i\bar{i};ab} a_a^\dagger a_b^\dagger + \sum_a c_{i\bar{i};a\bar{i}} a_a^\dagger a_{\bar{i}}^\dagger\right)|\theta\rangle$$

$$= \prod_{i=1}^{N/2}\left(1 + \sum_{\substack{a\in S_1, b\in S_2 \\ S_1\cap S_2=\varnothing}} c_{i\bar{i};ab} a_a^\dagger a_b^\dagger a_{\bar{i}} a_i + \sum_a c_{i\bar{i};a\bar{i}} a_a^\dagger a_{\bar{i}}^\dagger a_{\bar{i}} a_i\right) a_i^\dagger a_{\bar{i}}^\dagger|\theta\rangle \quad (18)$$

$$= \prod_{i=1}^{N/2}\left(1 + \sum_{\substack{a\in S_1, b\in S_2 \\ S_1\cap S_2=\varnothing}} c_{i\bar{i};ab} a_a^\dagger a_b^\dagger a_{\bar{i}} a_i + \sum_a c_{i\bar{i};a\bar{i}} a_a^\dagger a_{\bar{i}}^\dagger a_{\bar{i}} a_i\right)|\Phi_0\rangle$$

$$= \exp\left(\sum_{\substack{i,a\in S_1, b\in S_2 \\ S_1\cap S_2=\varnothing}} c_{i\bar{i};ab} a_a^\dagger a_b^\dagger a_{\bar{i}} a_i + \sum_{i,a} c_{i\bar{i};a\bar{i}} a_a^\dagger a_{\bar{i}}^\dagger a_{\bar{i}} a_i\right)|\Phi_0\rangle$$

In both cases, it is easy to show that analogues of Eq. (14) hold. Note that in the usual case in APsetG in which we divide the spin-orbitals according to their spin, APset1roGSD will preserve the $M_z$ value of the reference. On the other hand, APG1roSD will break this symmetry. Regarding their CC notation, APset1roGSD and APG1roSD could more succinctly be called sCCSD and gCCSD, respectively.

2.3 Breaking seniority

Taking another look at the operator $\hat{t}_{i\bar{i}}^{a\bar{i}} = a_a^\dagger a_{\bar{i}}^\dagger a_{\bar{i}} a_i$, we see that it can be rewritten as $\hat{t}_{i\bar{i}}^{a\bar{i}} = \hat{t}_i^a \hat{n}_{\bar{i}} = a_a^\dagger a_i \hat{n}_{\bar{i}}$. That is, as the multiplication of the traditional T$_1$ operator $\hat{t}_i^a$, and the occupation-number operator, $\hat{n}_{\bar{i}}$, corresponding to the spin-orbital paired with $i$. Hence, the quasi-T$_1$ operators introduced in the previous section will only act when they excite from a doubly occupied spatial orbital. Another way of saying this is that the wavefunctions discussed in 2.2 have single excitations that must increase the seniority[114-119] of the occupied manifold. For this reason, moving forward, the wavefunctions introduced in the previous section will be appended by the "sen-o" indicator.

This insight motivated us to look for other ways in which one could use the $T_1$ operators to break the seniority in different ways. Besides the previously discussed sen-o flavor, one can think of other three possibilities (all of these variants are illustrated in Fig. 2):

a) Breaking the seniority in the virtual space, by not allowing the S excitations to form doubly occupied pairs in virtual spatial orbitals: which we call sen-v.

b) Breaking the seniority in the occupied **and** in the virtual space. So, the S excitation must break a doubly occupied orbital and not form a pair in a virtual orbital: which we call sen-ov.

c) Using a $T_1$ operator with no restrictions: which we call sen-free.

| Ref. state | Seniority-breaking conditions for singles-like excitations | | | |
|---|---|---|---|---|
| HF | sen-o | sen-v | sen-ov | sen-free |
| | allowed / not allowed | allowed / not allowed | not allowed / not allowed | allowed / allowed |
| **Operators** | $a_a^\dagger a_k \hat{n}_{\bar{k}}$ | $a_a^\dagger a_k (1 - \hat{n}_{\bar{a}})$ | $a_a^\dagger a_k (1 - \hat{n}_{\bar{a}}) \hat{n}_{\bar{k}}$ | $a_a^\dagger a_k$ |

**Figure 2.** Diagrammatic and operator representation of allowed and not-allowed configurations of orbital levels in seniority-breaking conditions sen-o, -v, -ov, -free.

For example, in the case of AP1roG, these wavefunctions will have the form:

$$\left| \Psi_{\text{AP1roGSD\_sen-v}} \right\rangle = \prod_{i=1}^{N/2} \left( 1 + \sum_a c_{i\bar{i};a\bar{a}} \hat{t}_{i\bar{i}}^{a\bar{a}} \right) \prod_{i=1}^{N/2} \left( 1 + \sum_a c_{i\bar{i};a\bar{i}} \hat{t}_i^a (1 - \hat{n}_{\bar{a}}) \right) \left| \Phi_0 \right\rangle \quad (19)$$

$$\left| \Psi_{\text{AP1roGSD\_sen-ov}} \right\rangle = \prod_{i=1}^{N/2} \left( 1 + \sum_a c_{i\bar{i};a\bar{a}} \hat{t}_{i\bar{i}}^{a\bar{a}} \right) \prod_{i=1}^{N/2} \left( 1 + \sum_a c_{i\bar{i};a\bar{i}} \hat{t}_i^a \hat{n}_{\bar{i}} (1 - \hat{n}_{\bar{a}}) \right) \left| \Phi_0 \right\rangle \quad (20)$$

$$\left| \Psi_{\text{AP1roGSD\_sen-free}} \right\rangle = \prod_{i=1}^{N/2} \left( 1 + \sum_a c_{i\bar{i};a\bar{a}} \hat{t}_{i\bar{i}}^{a\bar{a}} \right) \prod_{i=1}^{N/2} \left( 1 + \sum_a c_{i\bar{i};a\bar{i}} \hat{t}_i^a \right) \left| \Phi_0 \right\rangle \quad (21)$$

with similar expressions holding for the APG1ro and APset1roG methods.

It is important to remark that, of all the ways of breaking seniority with single excitations, only sen-o leads to geminal wavefunctions. The other ways of increasing seniority (cf. Eqs. (19)-(21)) can only be expressed in CC form.

## 3. MODEL SYSTEMS AND COMPUTATIONAL DETAILS

### 3.1 Model systems

We assessed the performance of the previously discussed wavefunctions over simple model systems with 4, 6, 8, and 10 electrons. The four-electron systems include planar (S4, H4, P4, and D4) and non-planar (T4 and V4) arrangements (see Figs. 3 and 4). Their corresponding geometries were taken from the work of Paldus *et al.*[120,121] See the Supplementary Information details on the conformations of all the $H_4$ model systems.

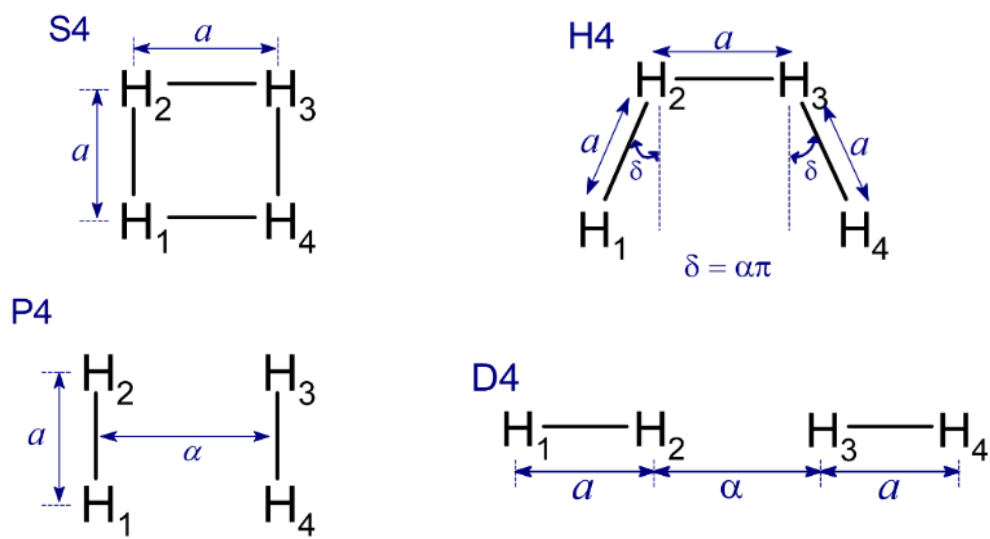

**Figure 3.** Nuclear configurations and definitions of the variable parameters for S4, H4, P4, and D4 models. The H atoms of $H_4$ cluster are labeled as 1, 2, 3 and 4 indicated in subscripts.

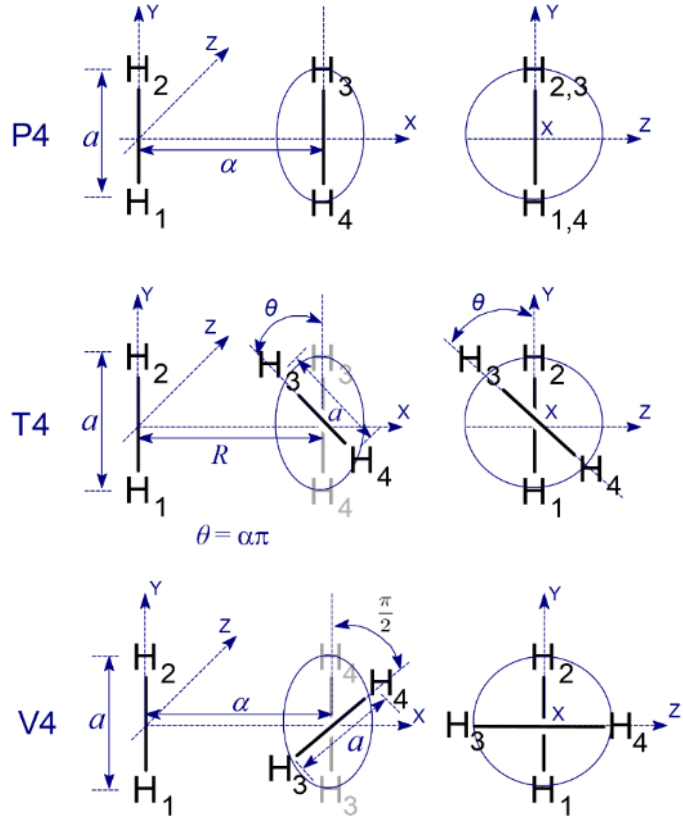

**Figure 4.** Nuclear configurations and definitions of the variable parameters for nonlinear P4, T4 and V4 models. The H atoms of $H_4$ cluster are labeled as 1, 2, 3 and 4 indicated in subscripts.

Additionally, we considered the venerable $BeH_2$ test case, which has been widely used as a toy model to examine the performance of strong-correlation methods along the insertion of Be into $H_2$. Because of the quasi-degeneracy of the $2s$ and $2p$ orbitals, $BeH_2$ exhibits both weak and strong correlation effects. We studied ten different geometries of this reaction following the previous study by Bartlett et al.[122] Here, each geometry is generated by varying the H-H bond length and perpendicular Be-$H_2$ bond distance. The cartesian coordinates of H atoms are listed in Table (1) considering Be atom at the origin. The energy values for $BeH_2$ geometries using 1-reference orbital geminal wavefunctions are plotted against the perpendicular distance between Be and $H_2$.

| Point | X | Y | Z |
| --- | --- | --- | --- |
| A | 0.0 | ± 2.54 | 0.0 |
| B | 0.0 | ± 2.08 | 1.0 |
| C | 0.0 | ± 1.62 | 2.0 |
| D | 0.0 | ± 1.39 | 2.5 |
| E | 0.0 | ± 1.275 | 2.75 |
| F | 0.0 | ± 1.16 | 3.0 |
| G | 0.0 | ± 0.93 | 3.5 |
| H | 0.0 | ± 0.70 | 4.0 |
| I | 0.0 | ± 0.70 | 6.0 |
| J | 0.0 | ± 0.70 | 20.0 |

**Table 1.** Cartesian coordinates (in a.u.) of H atoms for geometries along $C_{2v}$ potential energy sampling path of perpendicular insertion of Be into $H_2$. For all geometries, the Be atom is at (0.0, 0.0, 0.0).

The eight-electron examples encompass two model systems consisting of eight equidistant H atoms arranged in a line and a cubic geometry. Different geometries of linear $H_8$ chain are generated by varying all internuclear distances of adjacent H atoms in the interval of [0.5, 4] Å. The geometries for this model systems are obtained from Ref. [123].

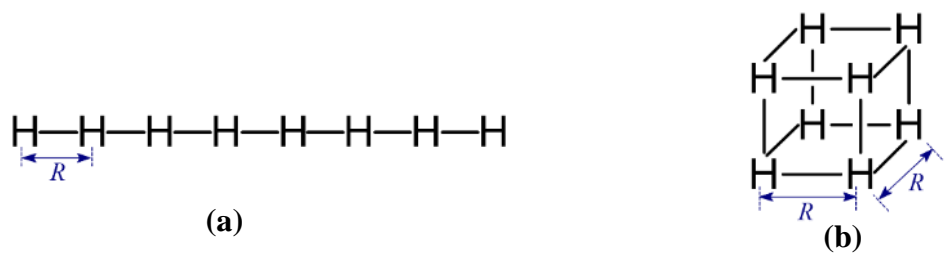

**Figure 5.** (a) Linear chain (1D) and (b) cube (3D) model of $H_8$ clusters. All geometries are parameterized by nearest-neighbor H-H bond distance, $R$.

Finally, four different model systems of 10 equally spaced H atoms proposed by Stair and Evangelista[124] were studied: chain (1D), ring (1D), sheet (2D), and pyramid (3D). These serve as a stringent test for strongly correlated methods, spanning different dimensionalities and non-innocent bond patterns.

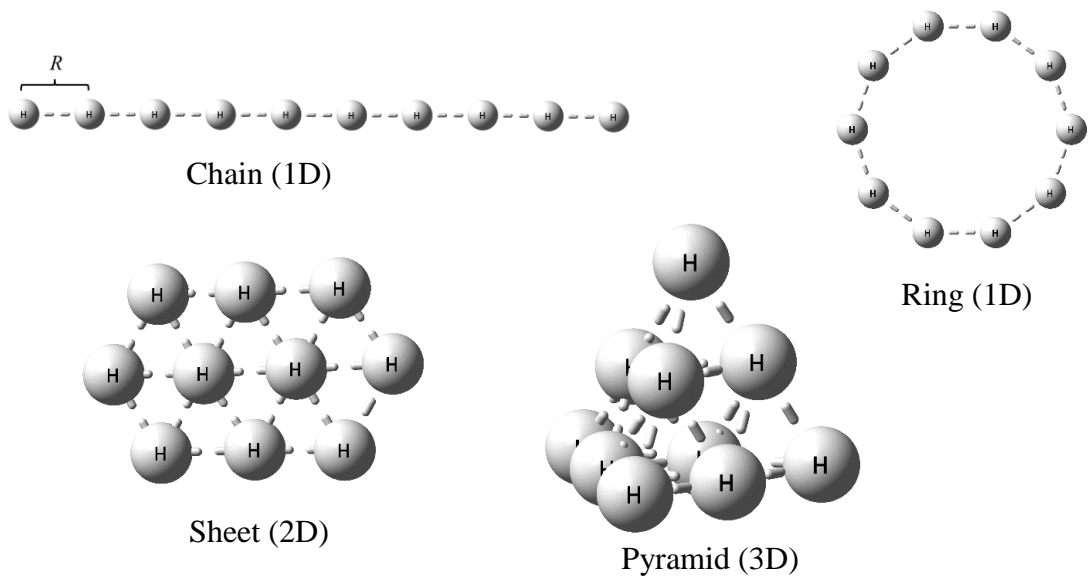

**Figure 6.** Ring, chain, sheet, and pyramid models of $H_{10}$ clusters. Geometries of all models are parameterized by the nearest-neighbor H-H bond distance ($R$).

3.2 Computational details

All the methods discussed here were implemented in a development branch of the Fanpy package,[125] using the FANCI framework[126] as a way to speed-up the implementation and testing of novel *ab initio* methods. All calculations used the STO-6G basis set. The Molecular Orbitals (MOs) were obtained from Restricted Hartree-Fock (RHF) calculations. One- and two-body integrals for the Hamiltonians were processed with HORTON.[127] The exact energy values for model systems of $H_4$, $H_8$ clusters and $BeH_2$ are obtained using the Full Configuration Interaction (FCI) method implemented in Fanpy. For the $H_{10}$ model systems, FCI energies are obtained from the work of Stair and Evangelista.[124]

## 4. COMPUTATIONAL RESULTS

4.1 1-reference orbital geminal wavefunctions with Doubles

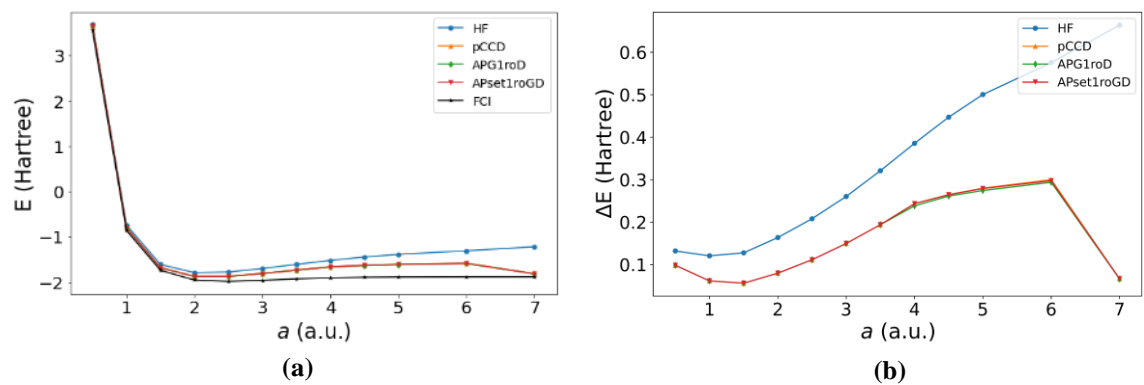

**Figure 7.** 1roD: (a) total energies and (b) energy errors ($\Delta E = E_{\text{wavefunction}} - E_{\text{FCI}}$). Square S4 model (STO-6G basis). Also shown, HF and FCI results.

In Figure 7, we show the ground state energies calculated using 1-reference orbital (1ro) geminal wavefunctions with only doubly excited configurations (1roD): AP1roG (shown as pCCD), APG1roD, and APset1roGD, for different geometries of S4 model parameterized by simultaneously varying all H-H bond distances. While the geminal wavefunctions obviously account for some of the missing correlation in HF, it is clear that without OO the geminal models do not significantly improve upon the uncorrelated results. All 1roD geminal wavefunctions have virtually the same energy values for $a \leq 3.5$ a.u.. Beyond this point, APG1roD (green curve) has the lowest energy among the three geminal wavefunctions. This is to be expected, given that APG1roD considers more orbital pairs compared to the other two geminal wavefunctions. The comparison with FCI, Fig. 7(b), shows the 1-reference orbital geminal wavefunctions with doubles have a maximum difference of around 0.3 Hartrees ($E_h$) at $a = 6$ a.u. for AP1roG. However, all geminal wavefunctions converge almost to the FCI energy value for the geometry with $a = 7$ a.u., with a minimum difference of 0.0661 $E_h$. The other $H_4$ model systems present a similar behavior (see the Supplementary Information, section S2.1, for detailed results). Overall, ignoring the OO step the 1roD wavefunctions have errors in the order of ~ 0.1 $E_h$ so, at best, they are qualitatively correct, but remain far from quantitatively accurate.

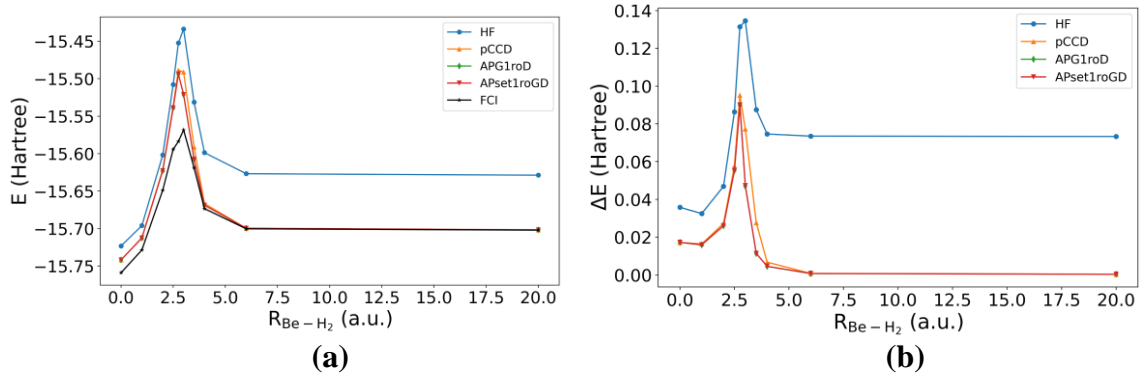

**Figure 8.** 1roD: (a) total energies and (b) energy errors ($\Delta E = E_{\text{wavefunction}} - E_{\text{FCI}}$), for $C_{2v}$ insertion of Be into $H_2$ (STO-6G basis). Also shown, HF and FCI results.

The $C_{2v}$ insertion of Be into $H_2$ presents a more challenging test, due to the multiple configurations of $BeH_2$ contributing significantly to total ground state energy calculations, thus making it difficult for single reference methods to describe this process. From Fig. 8(a), we see that 1roD wavefunctions, especially APset1roGD and APG1roD, follow the same behavior as the reference FCI results at very close and very far internuclear separations. However, for intermediate geometries with $R_{\text{Be-H}_2}$ = 2.5, 2.75, 3.0 a.u., the energy error is substantially higher, once again approaching the 0.1 $E_h$ region, as in the $H_4$ case. Notice how FCI finds a maximum at $R_{\text{Be-H}_2}$ = 3.0 a.u. whereas the 1roD wavefunctions reach their maximum a bit earlier, at 2.75 a.u., thus even failing to locate the "transition state" of this process. These disagreements eventually die-off at larger separations, with errors now around 0.355 m$E_h$.

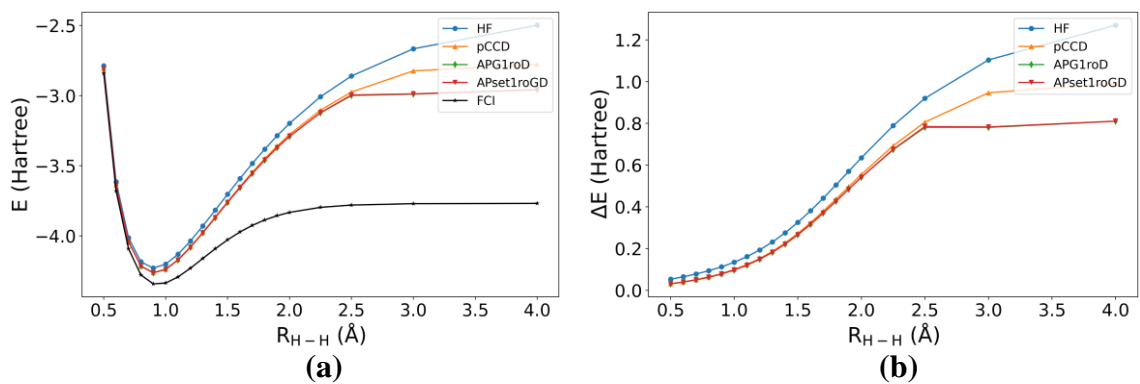

**Figure 9.** 1roD: (a) total energies and (b) energy errors ($\Delta E = E_{\text{wavefunction}} - E_{\text{FCI}}$). Linear $H_8$ model (STO-6G basis). Also shown, HF and FCI results.

Fig. 9 shows the 1roD, HF, and FCI results for the symmetric dissociation of linear $H_8$. For H-H separations beyond 0.8 Å, the geminal wavefunctions are essentially parallel to the HF results, leading to a very big gap with respect to FCI. Particularly problematic is the behavior of APG1roD and APset1roGD when $R_{H-H} > 2.5$ Å. These wavefunctions did not converge properly in this regime, showing a plateau of energy values corresponding to errors almost on the order of 1 $E_h$. pCCD tends to follow the HF results more closely at large separations, so it also presents very large errors. Once again, the pivotal role of the OO becomes clear, since without the correct orbitals we cannot even obtain size consistent results and properly describe the dissociation limit for a system as simple as $H_8$. (Corresponding results for cubic $H_8$ are presented in the Supplementary Information, section S2.1.3.)

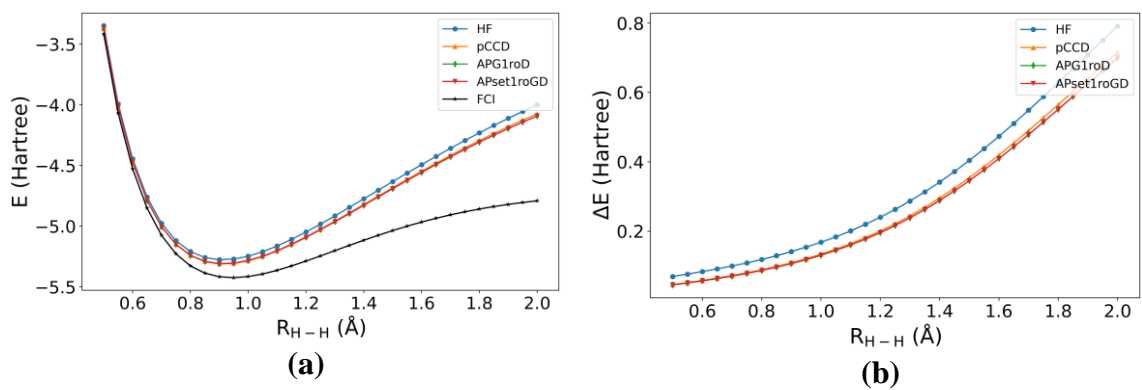

**Figure 10.** 1roD: (a) total energies and (b) energy errors ($\Delta E = E_{\text{wavefunction}} - E_{\text{FCI}}$). Linear $H_{10}$ model (STO-6G basis). Also shown, HF and FCI results.

As expected, the $H_{10}$ clusters show even more pathological behavior in the absence of orbital optimization (full details are in the Supplementary Information, section S2.1.4). From the simplest 1D system (Fig. 10) to the 3D pyramid (Fig. 11), the geminal wavefunctions once again tend to follow HF more closely than they follow FCI, totally failing to correctly dissociate. The geminal wavefunctions without OO perform rather poorly even at the minimum of the dissociation curve, with errors already in the order of 0.1 $E_h$ for linear $H_{10}$. The behavior at shorter $R_{H-H}$ is not as bad for pyramid $H_{10}$, but it nonetheless quickly shows signs of divergence with errors going over 0.5 $E_h$ over 1.8 Å.

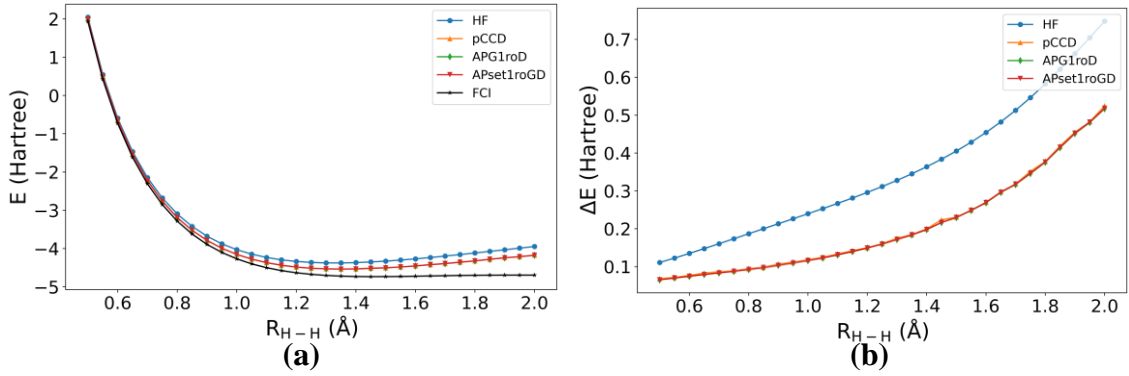

**Figure 11.** 1roD: (a) total energies and (b) energy errors ($\Delta E = E_{\text{wavefunction}} - E_{\text{FCI}}$). Pyramid $H_{10}$ model (STO-6G basis). Also shown, HF and FCI results.

4.2 1-reference orbital geminal wavefunctions with Singles and Doubles: seniority occupied restriction

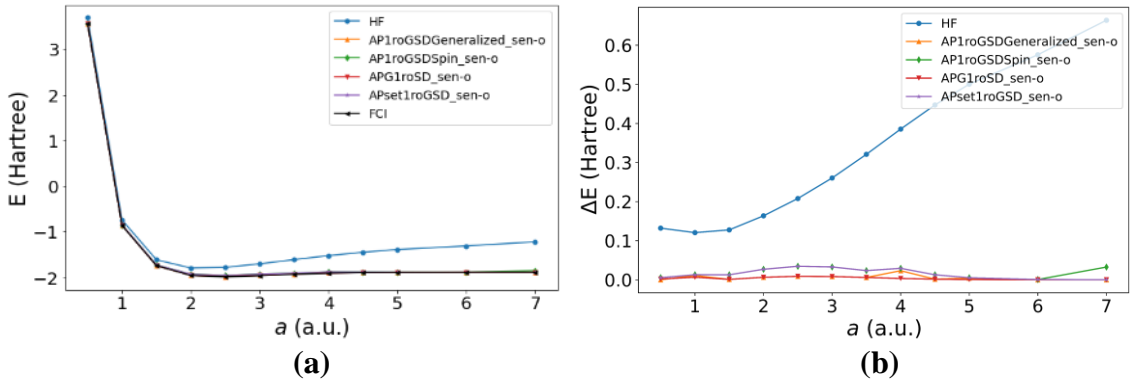

**Figure 12.** 1roSD_sen-o: (a) total energies and (b) energy errors ($\Delta E = E_{\text{wavefunction}} - E_{\text{FCI}}$). Square S4 model (STO-6G basis). Also shown, HF and FCI results.

The inclusion of single excitations in the 1-reference orbital single and double geminal wavefunctions with seniority restrictions in the occupied space (1roSD_sen-o) makes an instant impact on the geminal results, both qualitatively and quantitatively. Remarkably, these results did not include any explicit and/or separate OO step, with the modified S operators effectively carrying out this job thanks to the inclusion of Thouless rotations. Even for the S4 $H_4$ model, the difference between Figs. 12(a) and 7(a) is already evident. For the most part, the 1roSD_sen-o wavefunctions are essentially overlapping with the FCI results. In particular, the APG1roSD_sen-o method performs remarkably well for all inter-nuclear distances. Even the less accurate methods in this case (the

APsetG and spin-restricted AP1roG flavors) still outperform their counterparts without the S excitations, having errors one order of magnitude smaller than the 1roD approaches.

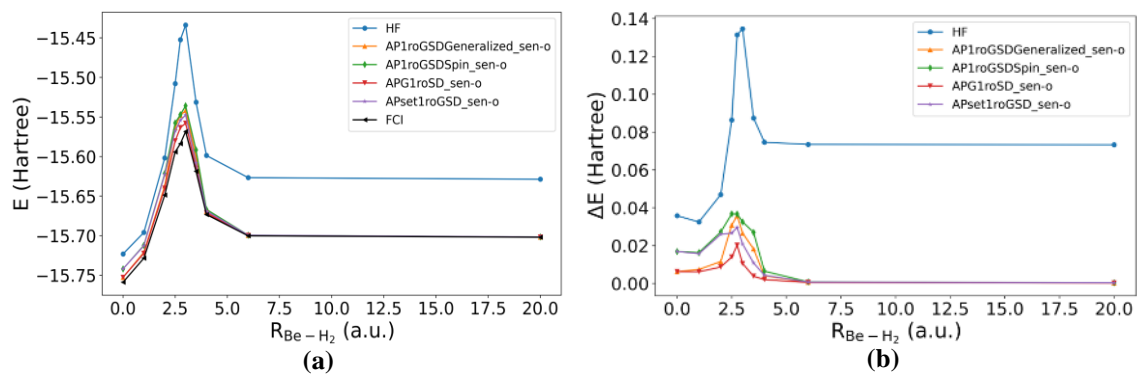

**Figure 13.** 1roSD_sen-o: (a) total energies and (b) energy errors ($\Delta E = E_{\text{wavefunction}} - E_{\text{FCI}}$) for $C_{2v}$ insertion of Be into $H_2$ (STO-6G basis). Also shown, HF and FCI results.

The description of the $C_{2v}$ insertion of Be into $H_2$ also shows great improvement, with all the 1roSD_sen-o methods now able to correctly identify the energy maximum at $R_{\text{Be-H}_2} = 3.0$ a. u. Unsurprisingly, APG1roSD_sen-o remains the most accurate method, with the other 1roSD_sen-o wavefunctions having similar performances in the region from 2.0 to 4.0 a. u.

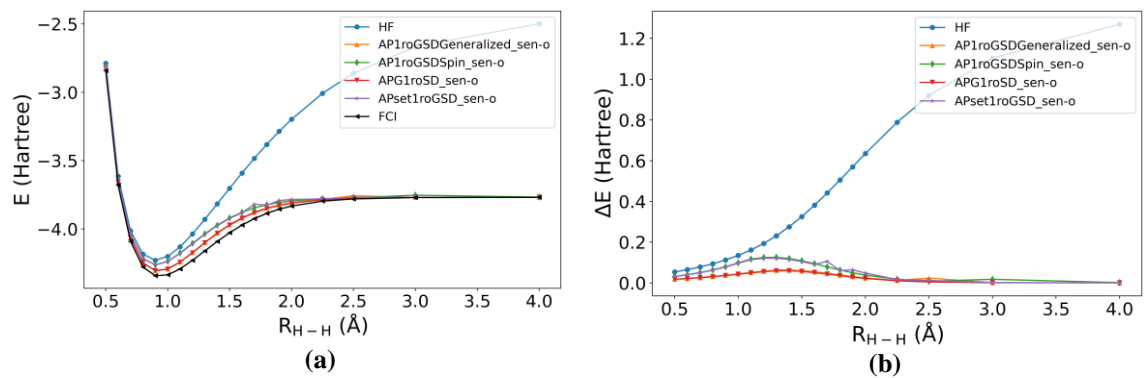

**Figure 14.** 1roSD_sen-o: (a) total energies and (b) energy errors ($\Delta E = E_{\text{wavefunction}} - E_{\text{FCI}}$). Linear $H_8$ model (STO-6G basis). Also shown, HF and FCI results.

In Fig. 9, we observed that 1roD geminals essentially diverged from FCI, having a behavior much closer to the HF results for the $H_8$ linear system. On the contrary (as shown in Fig. 14), the 1roSD methods give the proper dissociation limit to eight H atoms, with all the geminal wavefunctions essentially overlapping with FCI for large $R_{\text{H-H}}$ values.

Moreover, the S correction also significantly improves the quantitative agreement. Now the 1roSD versions of APG and (generalized) AP1roG have maximum errors ~ 60 m$E_h$, with APsetG and (spin-restricted) AP1roG ~ 120 m$E_h$. Notably, the biggest discrepancies are now seen in the region close to the equilibrium distance, so we can anticipate that these errors are due to the deficient treatment of dynamic correlation in these new wavefunctions

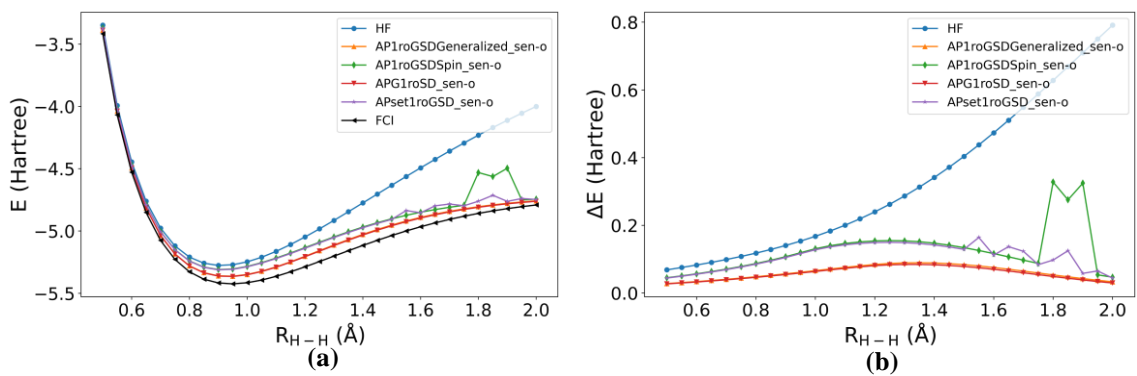

**Figure 15.** 1roSD_sen-o: (a) total energies and (b) energy errors ($\Delta E = E_{\text{wavefunction}} - E_{\text{FCI}}$). Linear H$_{10}$ (STO-6G basis). Also shown, HF and FCI results.

Properly describing the H$_{10}$ clusters is the most demanding test for the new wavefunctions. Fortunately, the inclusion of the S operators now allows for a proper description of the dissociation limit in the majority of cases. The 1roSD flavors of APG and (generalized) AP1roG are almost indistinguishable for the 1D systems. However, as we move to 2D and (especially) 3D, AP1roG starts to deteriorate but, reassuringly, APG1roSD_sen-o remains a very robust method. Following a trend observed in the other model systems, APsetG and (spin-restricted) AP1roG can be sometimes difficult to converge (see, for instance, the kinks in the AP1roGSDspin_sen-o results in Fig. 15) or show a problematic behavior at dissociation (Fig. 16). This shows the advantage of breaking the $M_z$ symmetry.

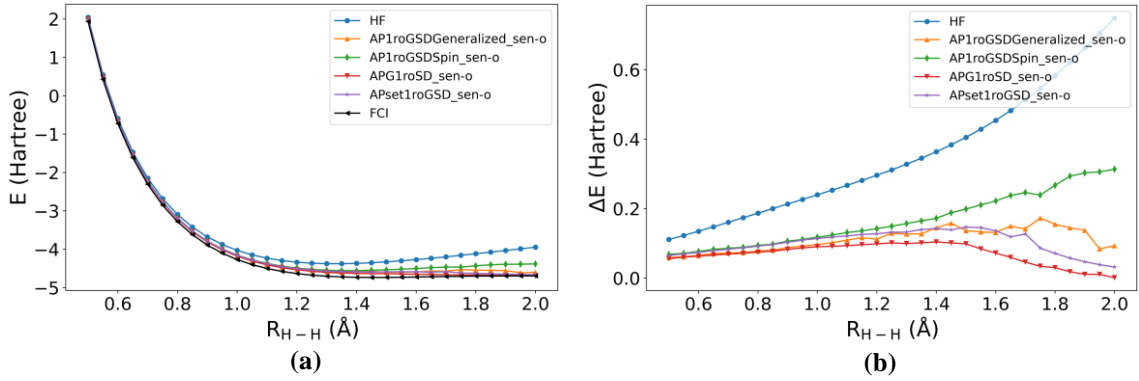

**Figure 16.** 1roSD_sen-o: (a) total energies and (b) energy errors ($\Delta E = E_{\text{wavefunction}} - E_{\text{FCI}}$). Pyramid $H_{10}$ (STO-6G basis). Also shown, HF and FCI results.

4.3 1-reference orbital geminal wavefunction with Singles and Doubles: other seniority patterns

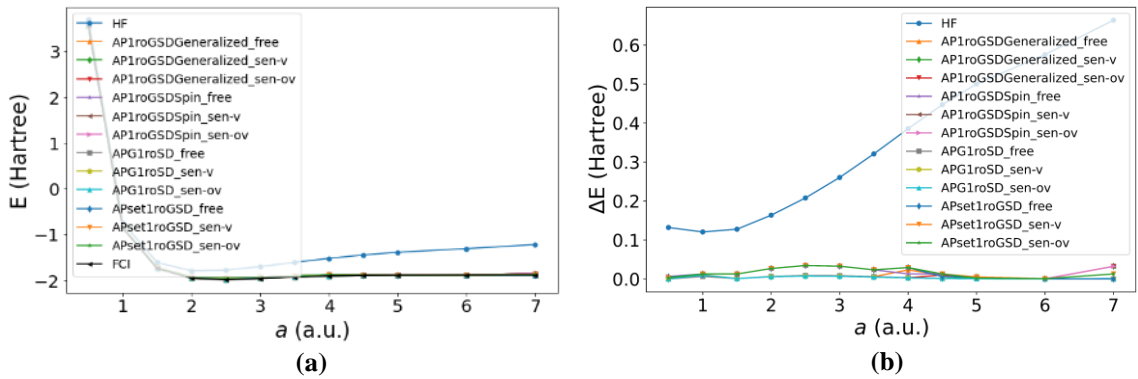

**Figure 17.** 1roSD_sen-ov, sen-v, and sen-free: (a) total energies and (b) energy errors ($\Delta E = E_{\text{wavefunction}} - E_{\text{FCI}}$). Square S4 model (STO-6G basis). Also shown, HF and FCI results.

In the previous section we observed that the addition of sen-o S improved the performance of 1roD geminal wavefunctions. A similar trend follows for the $H_4$ systems when we explore the other seniority restrictions: sen-ov, sen-v, and sen-free (Fig. 33). These wavefunctions (which cannot be formulated using geminals) seem to behave like their sen-o counterparts, with the APG flavors once again proving to be the more accurate across inter-nuclear separations (followed closely by the generalized AP1roG variants).

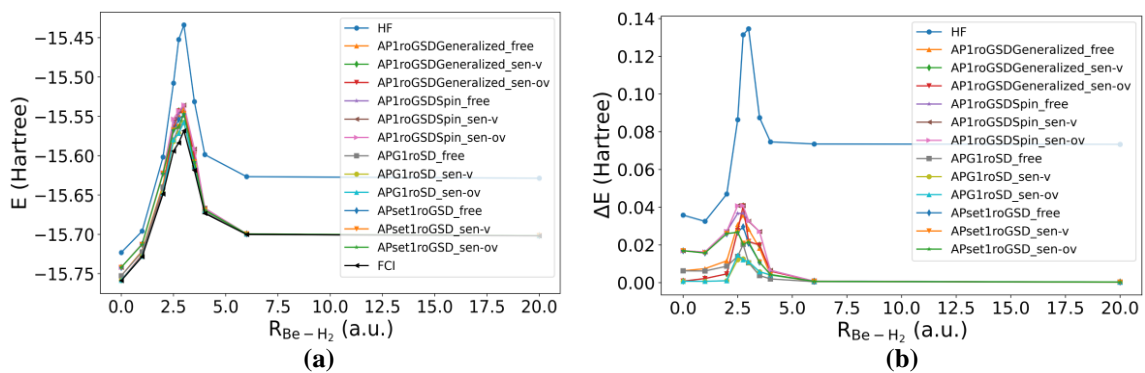

**Figure 18.** 1roSD_sen-ov, sen-v, and sen-free: (a) total energies and (b) energy errors ($\Delta E = E_{\text{wavefunction}} - E_{\text{FCI}}$) for $C_{2v}$ insertion of Be into $H_2$ (STO-6G basis). Also shown, HF and FCI results.

The non-geminal wavefunctions do not substantially improve upon the 1roSD_sen-o methods for $BeH_2$. All the nice properties observed in the previous section appear here as well (correct identification of the maximum along the dissociation curve, and perfect dissociation limit), while showing errors of ~ 0.01 $E_h$ in the region around $R_{\text{Be-H}_2} = 2.5$ a.u.

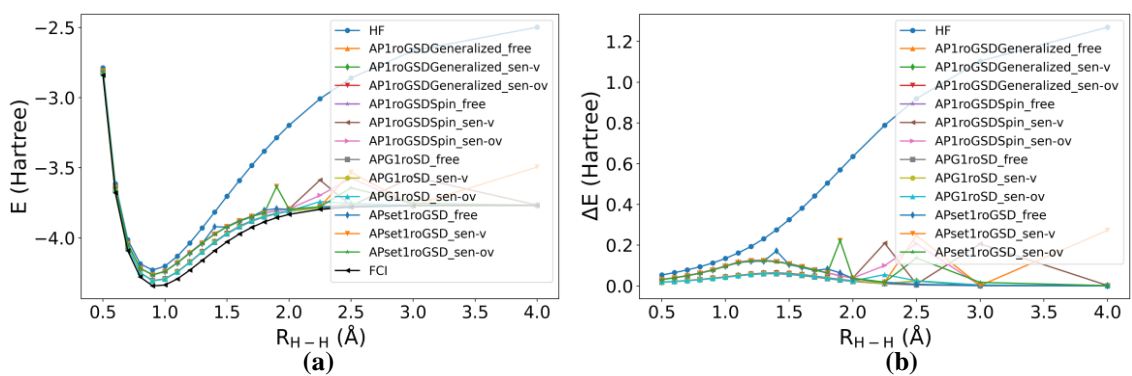

**Figure 19.** 1roSD_sen-ov, sen-v, and sen-free: (a) total energies and (b) energy errors ($\Delta E = E_{\text{wavefunction}} - E_{\text{FCI}}$). Linear $H_8$ model (STO-6G basis). Also shown, HF and FCI results.

When we increase the number of electrons we start seeing some deviations among the seniority-restricted flavors. For instance, while all sen-o wavefunctions gave smooth dissociation curves for linear $H_8$, the non-geminal variants (Fig. 19) are more pathological. With the notable exception of APG (and, in many cases, AP1roG), the other geminal-inspired wavefunctions proved very numerical unstable, resulting in pronounced kinks in the potential energy curve

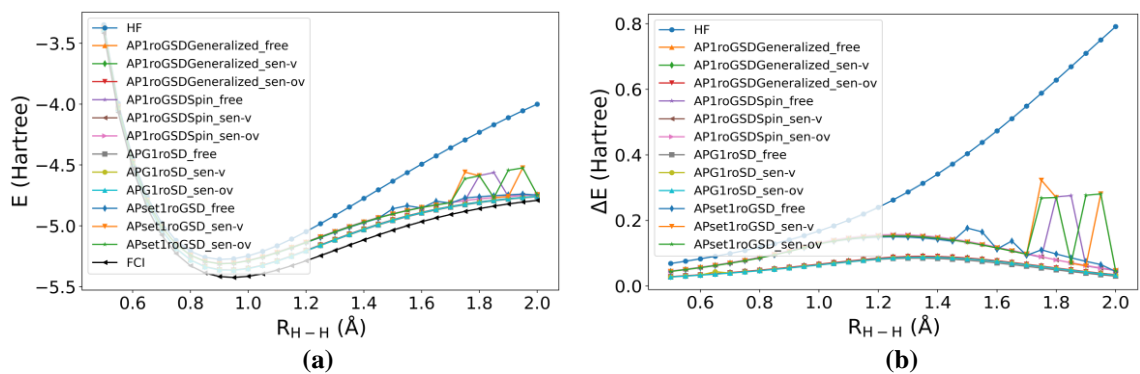

**Figure 20.** 1roSD_sen-ov, sen-v, and sen-free: (a) total energies and (b) energy errors ($\Delta E = E_{\text{wavefunction}} - E_{\text{FCI}}$). Linear $H_{10}$ (STO-6G basis). Also shown, HF and FCI results.

Unsurprisingly, $H_{10}$ continues the trend observed in $H_8$ of increased instability at larger H-H separation. This behavior (already problematic in the one-dimensional (linear) $H_{10}$ chain, Fig. 20) is even more pronounced in the 3D case (Fig. 21). Overall, the APG variants are the most well-behaved, with the generalized AP1roG flavors resulting also in relatively small errors, but with dissociation curves that are not as smooth.

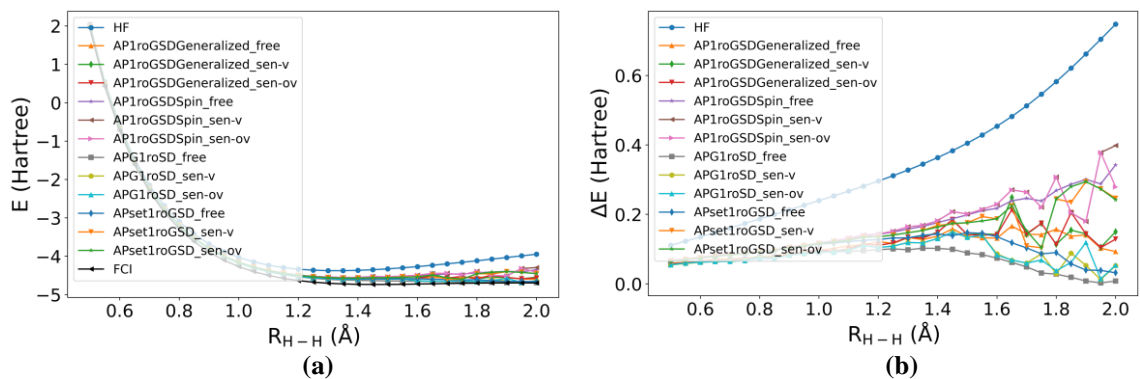

**Figure 21.** 1roSD_sen-ov, sen-v, and sen-free: (a) total energies and (b) energy errors ($\Delta E = E_{\text{wavefunction}} - E_{\text{FCI}}$). Pyramid $H_{10}$ (STO-6G basis). Also shown, HF and FCI results.

## 5. CONCLUSIONS

We delved deeper into the connection between geminal and CC wavefunctions. Building from the AP1roG/pCCD equivalency, we showed that the 1-reference orbital construction can be equally applied to more general geminal wavefunctions, like APsetG and APG. This allows us to approximate these intractable geminal wavefunctions with tractable CC wavefunctions that yet still can be formulated as a product of quasiparticles. Essentially, these new wavefunctions are approximations to CCD, but they do not completely collapse when breaking several chemical bonds. The main problem with these new methods is that, like AP1roG/pCCD, they require an OO step to carefully establish the correct pairing between the rotated molecular orbitals. As shown above, failure to include the OO leads to calculations that fail even qualitatively, being unable to properly dissociate even a few atoms.

As a way to bypass the need to perform the OO, we showed how one can include single-like effects in the geminal framework. Specifically, single excitations can masquerade as double excitations, thereby retaining two-particle structure of geminals while gaining the flexibility to let the wavefunction "adjust" the MOs via Thouless rotations. This provides a black-box mechanism that gives qualitatively and quantitatively correct results. This "faux singles" correction recovers proper dissociation limits in all cases and captures a larger fraction of the correlation energy. Our tests on the more demanding model systems showed that breaking the $M_z$ symmetry not only led to more accurate results, but also to methods that were more computationally robust. The best example of this is the comparison between APset1roGSD_sen-o and AP1roGSDgeneralized_sen-o. Even though APset1roGSD_sen-o has a more general pairing pattern, A1roGSDgeneralized_sen-o (with substantially fewer D operators) routinely provides more accurate and consistent results.

We noted that the inclusion of S operators demanded the inclusion of particle-number operators that effectively constrained their application in order to remain within the geminal picture. This means that the S excitations are only allowed to act if they are breaking an occupied electron pair. That is, if they can only increase the seniority starting from the occupied MOs (hence why the "sen-o" suffix added to these wavefunctions). Motivated by this, we also explored other ways of incorporating S-like excitations, with different restrictions (sen-v, sen-ov) or without any restrictions altogether (sen-free). These other methods, however, do not correspond to products of geminals, being closer to "product CC" wavefunctions. This generalization did not affect the performance of the APG-related flavors, but it introduced significant challenges for the other geminal families. Even systems like the $H_8$ clusters, which could be easily converged with the sen-o methods, became highly difficult to optimize. This strongly suggests that keeping the geminal structure facilitates the treatment of strong correlation, in particular, for relatively "simple" wavefunctions, like those derived from APIG and APsetG.

The new methods discussed here, in particular, the geminal flavors that can be expressed as CC wavefunctions including S-like operators, show great promise in describing strongly correlated systems. More importantly, the recipes outlined here to turn seemingly intractable geminal methods into computationally feasible approaches (without having to perform additional OO steps) can be applied to explore other electron-pairing schemes. There remain, however, some important questions around this framework, being perhaps the most important how to include the missing dynamic correlation in this formalism. Additionally, it will be desirable to explore ways to alleviate the convergence issues of the $M_z$-preserving methods. These problems will be explored in upcoming contributions.

**Acknowledgements:** RAMQ, PBG, RAL, and TDK thank support from ORAU in the form of a Ralph E. Powe award. RAMQ and PBG thank MolSSI for a Software Fellowship. PWA thanks the Canada Research Chairs, NSERC, and the Digital Research Alliance of Canada. PWA acknowledges NSERC, the Canada Research Chairs, and the Digital Research Alliance of Canada.

# Supplementary Information
# Coupled Cluster-Inspired Geminal Wavefunctions

Pratiksha B. Gaikwad,[1] Taewon D. Kim,[1] M. Richer,[2] Rugwed A. Lokhande,[1] Gabriela Sánchez-Díaz,[2] Peter A. Limacher,[2] Paul W. Ayers,[2*] Ramón Alain Miranda-Quintana[1*]

1. Department of Chemistry and Quantum Theory Project, University of Florida, Gainesville, FL 32603, USA
2. Department of Chemistry and Chemical Biology, McMaster University, Hamilton, Ontario, L8S 4M1, Canada

Emails: quintana@chem.ufl.edu, ayers@mcmaster.ca

**S1. MODEL SYSTEMS**

We assessed the performance of the previously discussed wavefunctions over simple model systems with 4, 6, 8, and 10 electrons. A comprehensive summary of the results can be found in: https://github.com/mqcomplab/GemCC_Data.

S1.1 Planar $H_4$ models

The geometries of the four-electron model systems with two interacting hydrogen molecules $(H_2)_2$ were taken from the work of Paldus *et al*.

S1.1.1 Trapezoidal model (H4)

The isosceles trapezoidal geometries are generated by breaking the H(1)-H(4) bond in the square arrangement of $(H_2)_2$ and by varying the bond angle $\delta = \angle H(1)$-$H(2)$-$H(3) - \pi/2 = \angle H(2)$-$H(3)$-$H(4) - \pi/2$. The "degeneracy parameter" $\alpha$ defined as $\alpha = \delta/\pi$ varies within $\alpha \in [0, \frac{1}{2}]$ as $\delta \in [0, \pi/2]$. Thus, $\alpha = 0$ gives the square confirmation whereas $\alpha = \frac{1}{2}$ results in a linear arrangement. Here, we study the H4 model for three different H-H bond lengths ($a = 1.2, 1.6, 2.0$ a.u.). For each of these $a$'s, we vary the $\delta$ values with different $\alpha$: $\alpha = [0.005, 0.01, 0.02, 0.05, 0.1, 0.015, 0.2, 0.25, 0.3, 0.4, 0.5]$.

S1.1.2 Rectangular model (P4)

In P4, two hydrogen molecules with same H-H bond distance H(1)-H(2) = $a$ = H(3)-H(4) are arranged in parallel with all the nuclei forming a rectangle. Varying the distance between two hydrogen molecules, given by $\alpha$, leads to different geometries of P4. Here, we examined both highly degenerate compressed geometries of P4 model ($\alpha < a$) and geometries in the dissociation limit ($\alpha \gg a$). Going from the S4 to the P4 model follows the consecutive breaking of two H-H bonds, leading to the dissociation of the H$_4$ cluster.

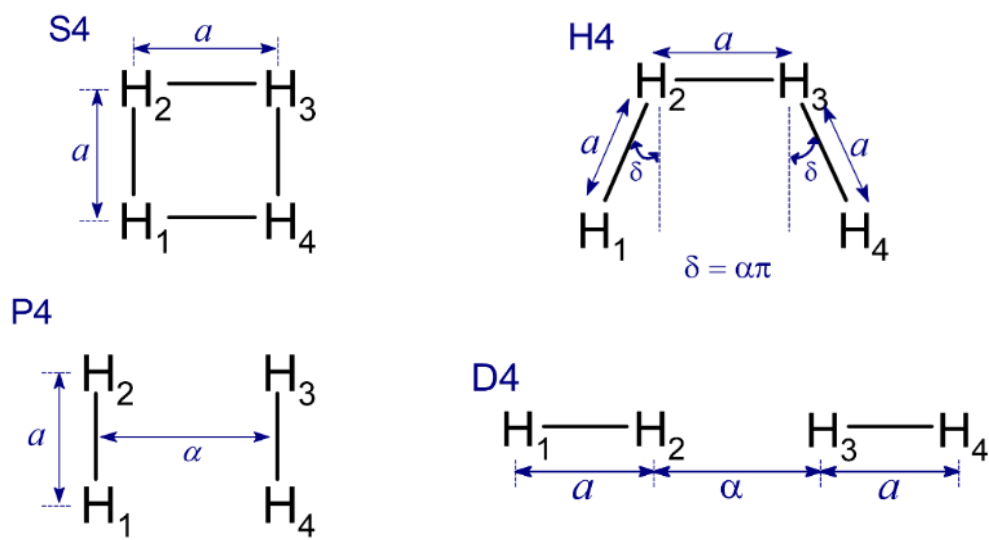

**Figure S1.** Nuclear configurations and definitions of the variable parameters for S4, H4, P4, and D4 models. The H atoms of H$_4$ cluster are labeled as 1, 2, 3 and 4 indicated in subscripts.

S1.1.3 Linear model (D4)

In the D4 model, two collinear hydrogen molecules of H-H bond distance $a$ are separated by a distance $\alpha$ distance apart from each other. Different geometries of the D4 model are generated with fixed $a$ = 2.0 a.u. varying for $\alpha < a$ and $\alpha \gg a$ region.

S1.2 Non-planar H$_4$ models

The non-planar H$_4$ models were obtained from Piecuch, P., & Paldus, J. (1994). T4 and V4 are generated from the previously discussed rectangular P4 model. As shown in Fig. S2, in the P4 model, two hydrogen molecules with same bond length are arranged in parallel along the Y-axis with H nuclei in a rectangular configuration.

S1.2.1 T4 model

The T4 model is obtained from the P4 model by rotating the H(3)-H(4) atoms around the center of the hydrogen molecule in the Y-Z plane by an angle $\theta$. Thus, the $\theta$ rotation is along the $C_2$ axis of symmetry (the X-axis). Different T4 geometries were generated by keeping the H-H bond distance same $a = 2.0$ a.u. for both $H_2$ molecules and varying the angle of rotation $\theta$ from 0 to $\theta = \pi/2$ while keeping the distance between two hydrogen molecules $R$ fixed. As indicated in the schematic presentation of the models, the angular parameter $\alpha$ is defined as $\alpha = \theta/\pi$, thus $\alpha = [0, \frac{1}{2}]$. Overall, we examined six sets of T4 variants corresponding to six different $R$ values. For each fixed intermolecular separation value $R$, energy values are plotted for varying $\theta$. For all each set of geometries, the bond length of both hydrogen molecules is fixed to $a = 2.0$ a.u..

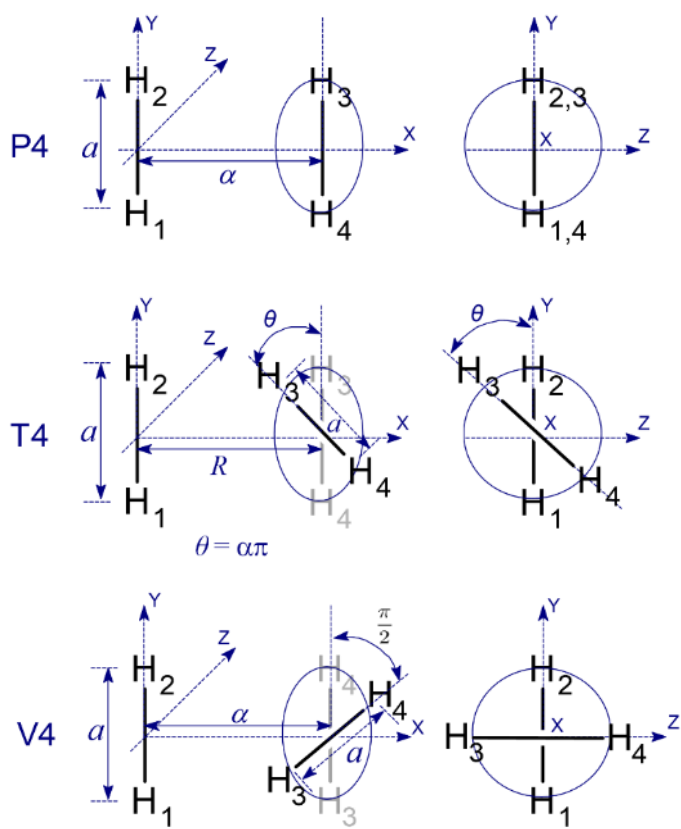

**Figure S2.** Nuclear configurations and definitions of the variable parameters for nonlinear P4, T4 and V4 models. The H atoms of $H_4$ cluster are labeled as 1, 2, 3 and 4 indicated in subscripts.

S1.2.2 V4 model

The T4 model with θ = π/2 leads to the V4 model, a maximally twisted T4 variant. So, both hydrogen molecules with same bond length, $a = 2.0$ a.u. become perpendicular to each other. The intermolecular distance for V4 model is shown by $\alpha$ which is analogous to $R$ from the T4 case. The V4 model is examined for different geometries generated by varying $\alpha$ from $\alpha < a$ to $\alpha \gg a$. With $\alpha = \sqrt{2} = a/\sqrt{2}$, H nuclei in the V4 model form a regular tetrahedral arrangement.

## S2. EXTENDED COMPUTATIONAL RESULTS

S2.1 1-reference orbital geminal wavefunctions with Doubles

S2.1.1 Planar H$_4$ models

S2.1.1.1 Trapezoidal model (H4)

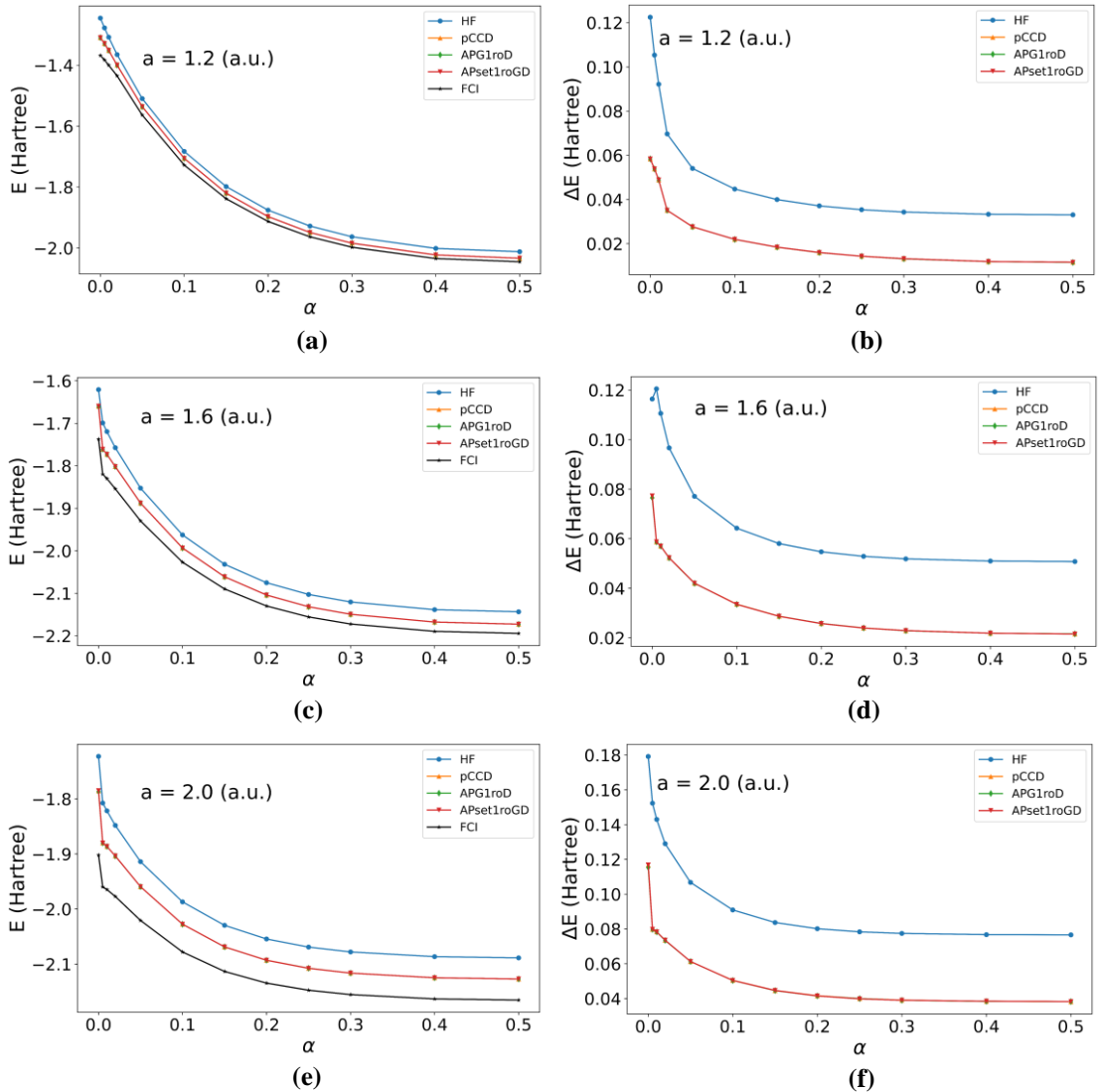

**Figure S3.** 1roD: (a) total energies and (b) energy errors ($\Delta E = E_{\text{wavefunction}} - E_{\text{FCI}}$). Trapezoidal H4 model (STO-6G basis). Also shown, HF and FCI results.

In Fig. 7, we observed the results for ground state energies using 1-reference orbital (1ro) geminal wavefunctions with only doubly excited configurations (1roD) start for the S4 square model. Breaking one H-H bond of S4 leads to the trapezoidal H4 model. Here, 1roD geminal wavefunctions energy values are presented for three different sets of H4 model, where each model has fixed H-H bond distance: (i) compressed ($a = 1.2$ a.u.) (ii) near-equilibrium ($a = 1.6$ a.u.) and (iii) stretched ($a = 2.0$ a.u.) geometries. For each set, energy values are presented by varying the degeneracy parameter $\alpha = \delta/\pi$ (section S1.1.1). The potential energy surface is obtained by varying $\alpha$ from $\alpha = 0$ (square arrangement) to $\alpha = 0.5$ (linear chain) involving the dissociation of a single H-H bond of S4 model. In Fig. S3, we see that the total ground state energy value decreases as $H_4$ goes from 2D trapezoid to 1D chain (H(1) and H(4) are pulled apart) by increasing the $\alpha$. Starting from the top plot for compressed geometries ($a = 1.2$ a.u.) to the bottom plot for stretched geometries ($a = 2.0$ a.u.) in the left column, the overall total energy value of 1roD wavefunctions decreases.

In Fig. S3 (a), (c), and (e), we see that all the 1roD energies are almost overlapping with each other. The total energy plot for H4 model with compressed geometries (Fig. S3 (a)) show a gradual decrease in energy with gradual initial increments in $\alpha$ values. But in Fig. S3 (b) and (c), we observe there is a drastic decrease in both the geminal wavefunctions and FCI energies with initial small increments in $\alpha$ values.

S2.1.1.2 Rectangular model (P4)

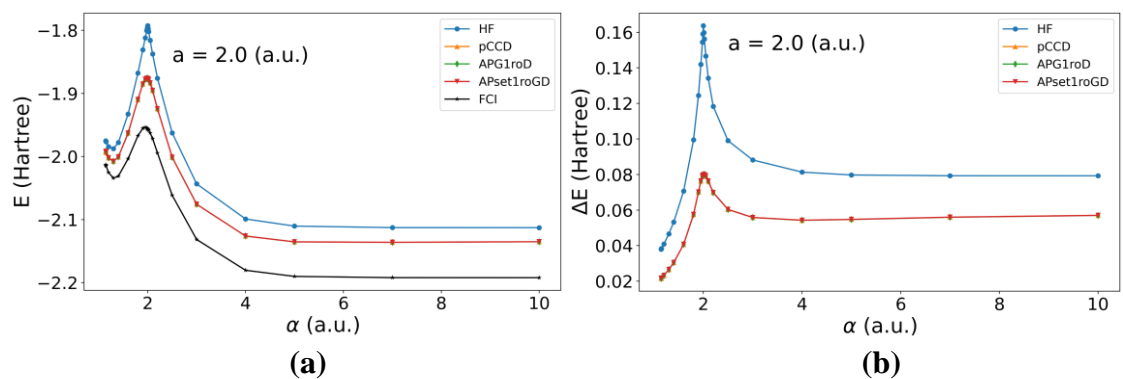

**Figure S4.** 1roD: (a) total energies and (b) energy errors ($\Delta E = E_{\text{wavefunction}} - E_{\text{FCI}}$). Rectangular P4 model (STO-6G basis). Also shown, HF and FCI results.

The rectangular arrangement of $H_4$ cluster is obtained by simultaneously breaking two H-H bonds of S4 model giving P4 model. We examined the P4 model by varying the intermolecular distance (shown by $\alpha$) between the centers of parallel H-H bonds of equal lengths ($a = 2.0$ a.u.) of the two $H_2$ molecules. Ground-state energy values of 1-reference orbital geminal wavefunctions are obtained for different geometries by non-linearly varying $\alpha$ in the interval of [$\sqrt{2}$, 10] a.u.. The total energy of the P4 model (Fig. S4) gradually decreases as the intermolecular distance increases from $\alpha = 1.1428$ a.u. to $\alpha = 1.3$ a.u., typical H-H bond length. Further, the total energy values of 1roD wavefunctions along with HF and FCI gradually increase with increasing $\alpha$ and reach their highest value when $\alpha = a = 2.0$ a.u., at which all four hydrogen atoms are arranged in a square form. For $\alpha > a$, the total energy values for all wavefunctions decrease further. For P4 model, all 1roD geminal wavefunctions have virtually the same energy values, similar to the results from S4 model, but still far from the FCI energies especially for the dissociated geometries of P4 model ($\alpha > a$). Without the orbital optimization (OO) the 1roD geminal wavefunctions have errors in the order of ~ 0.01 $E_{\text{h}}$.

S2.1.1.3 Linear model (D4)

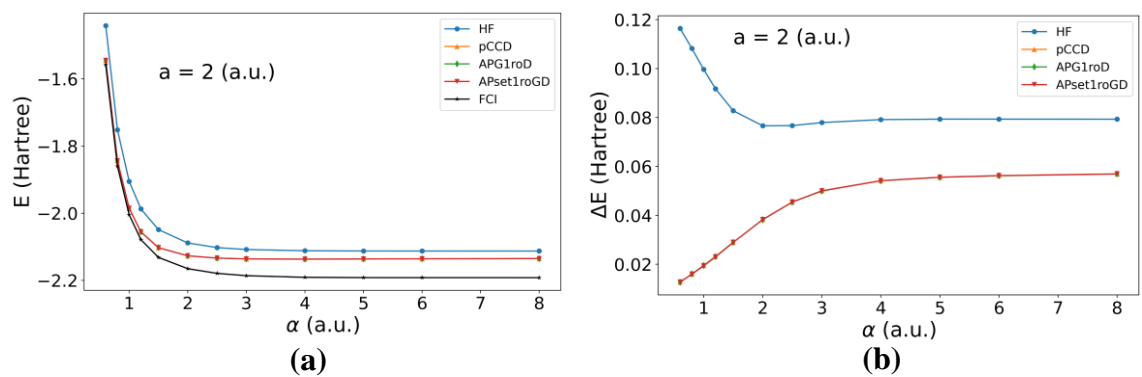

**Figure S5.** 1roD: (a) total energies and (b) energy errors ($\Delta E = E_{\text{wavefunction}} - E_{\text{FCI}}$). Linear D4 model (STO-6G basis). Also shown, HF and FCI results.

We obtained the potential energy surface of the linear model of 4 H atoms using 1roD geminal wavefunctions. The geometries of D4 model are parameterized by varying the distance between two hydrogen molecules of same bond length $a = 2.0$ a.u. arranged in a chain. The ground state total energies using 1roD (Fig. S5) show a good agreement with FCI energies for geometries of D4 with $\alpha \leq a$. Without the OO stop, 1roD results show a deviation from FCI following HF results for D4 geometries with $\alpha > a$. This shows that, due to the lack of OO step 1roD results qualitatively improve upon the uncorrelated HF energies but are quantitatively inadequate.

S2.1.2 Nonplanar H$_4$ models

S2.1.2.1 T4 model

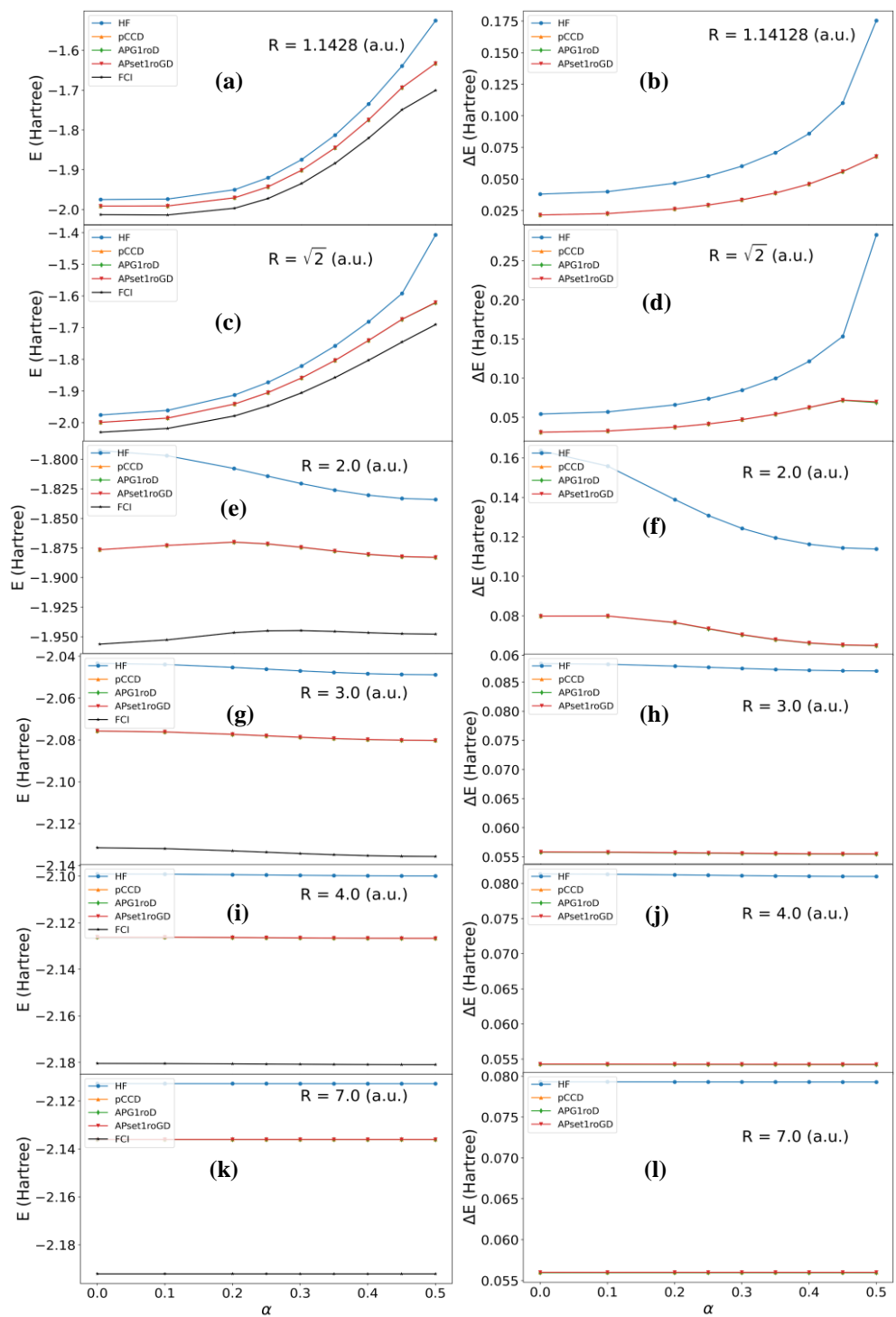

**Figure S6.** 1roD: (a) total energies and (b) energy errors ($\Delta E = E_{\text{wavefunction}} - E_{\text{FCI}}$). Nonlinear T4 model (STO-6G basis). Also shown, HF and FCI results.

The T4 model is obtained from the rectangular P4 model by rotating one molecular hydrogen bond around the axis passing through the center of both the hydrogen atoms. This allows us to study interactions between two hydrogen molecules in 3D. Different geometries of the T4 model are generated by varying the angle of rotation $\theta \in$

[0, $\pi/2$], so do the angular parameter $\alpha = \theta/\pi$, by keeping both H-H bond lengths $a = 2.0$ a.u. and intermolecular separation $R$, constant. As referred in previous studies, slightly stretched dihydrogen bonds are assumed to increase the quasidegeneracy effects. We have examined six different sets of geometries of T4 models for six different $R$ values while parameterizing the geometries in each set by single variable $\alpha$. The variation in $\alpha$ allowed us to study different conformations starting from planar ($\alpha = 0$) to the highly twisted form ($\alpha = \pi/2$) which leads to the potential energy surface (Fig. S6) corresponding the simultaneous stretching of H(1)-H(4) and H(2)-H(3) bonds.

Fig. S6(a)-(f) show that, total ground state energy for T4 models with $R < a$ drastically increase with increasing $\alpha$, for all 1roD wavefunctions. In this case, the energy graph for 1roD results show a maximum difference of 0.072 $E_h$ with FCI. For $R = a$, the total energy gradually converges towards FCI with error of $\Delta E_{min} \sim 0.065$ $E_h$ and for $R \gg a$, the total energy almost remains constant with increasing twist in the molecular conformation. In all the cases, 1roD geminal wavefunctions have energy errors $\sim 0.05$ $E_h$ for T4 models with $R > 3.0$. The overall 1-reference orbital geminal doubles results give a promise of better capturing of electronic correlation than HF but still having some gap in energy with respect to FCI.

S2.1.2.2 V4 model

In V4 model, we study the maximally twisted T4 model ($\theta = \pi/2$), in which two hydrogen molecules are perpendicular to each other at a distance of $\alpha$ (analogues to the variable $R$ from the T4 model). Different geometries of V4 model are studied by varying the internuclear separation $\alpha$ (shown along the X-axis in Fig. S2) in the interval of [1.2, 10] a.u.. The total ground state FCI energy (Fig. S7 (a)) increases initially with increasing $\alpha$, and it has a maximum for $\alpha = \sqrt{2}$ a.u. where, H(1)-H(4) = H(2)-H(3) = $a$ = 2.0 a.u.. The 1roD geminal wavefunctions reach their maximum a bit earlier at $\alpha = 1.3$ a.u., then energy values decrease gradually for $1.3 < \alpha \leq 4.0$ a.u.. For conformations of V4 model with $\alpha > 4.0$ a.u., 1roD deviate away from FCI. Overall, 1roD wavefunctions give qualitatively good results, but quantitatively far from the exact solution.

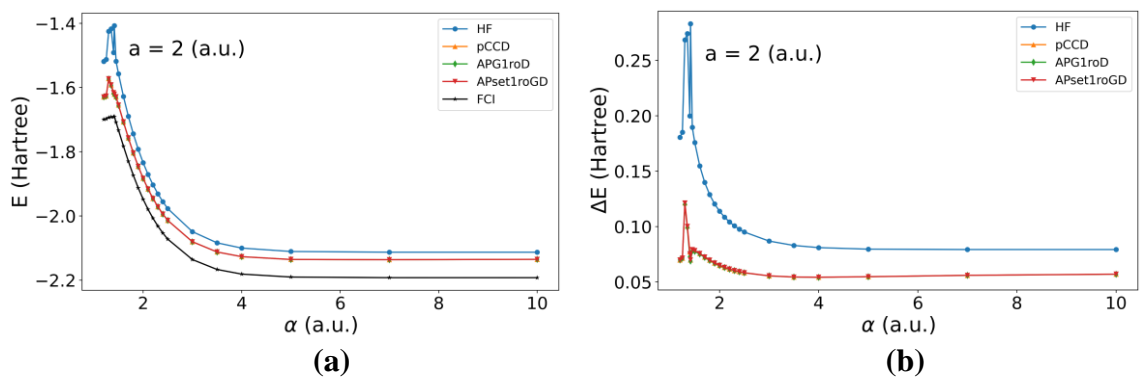

**Figure S7.** 1roD: (a) total energies and (b) energy errors ($\Delta E = E_{\text{wavefunction}} - E_{\text{FCI}}$). Nonlinear V4 model (STO-6G basis). Also shown, HF and FCI results.

### S2.1.3 $H_8$ Cube

The cubic model allows us to study the more complex geometry of the $H_8$ cluster compared to linear chain. In the case of chain model, we studied simultaneous stretching of seven H-H bonds, here in case of cube it will be twelve H-H bonds in 3D. In the case of cube (Fig. S8), the ground state energy values using 1roD wavefunctions exhibit a similar trend as in case of $H_8$ chain model (Fig. 9), except here the curves are not so smooth. These discrepancies address the lack of the OO step in 1roD geminal wavefunctions inadequately capturing the simultaneous breaking of multiple bonds 3D structure.

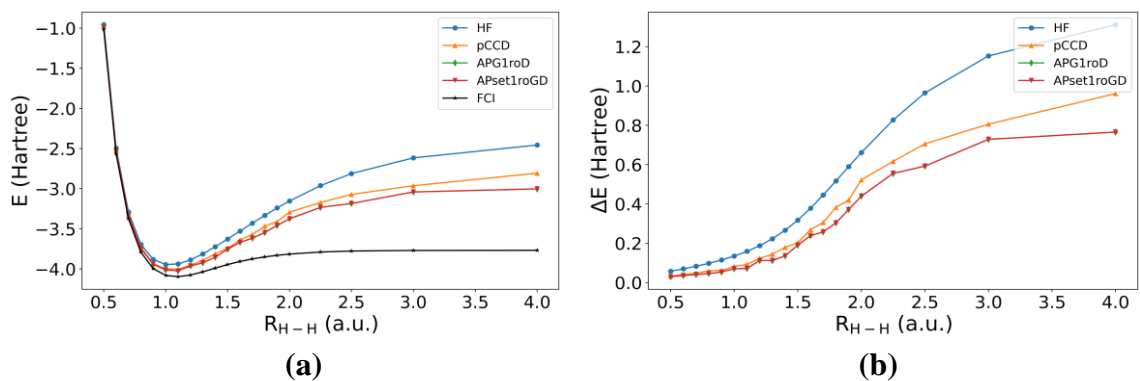

**Figure S8.** 1roD: (a) total energies and (b) energy errors ($\Delta E = E_{\text{wavefunction}} - E_{\text{FCI}}$). Cube $H_8$ model (STO-6G basis). Also shown, HF and FCI results.

### S2.1.4 $H_{10}$ Cluster

For comparison purposes, the FCI data and geometries of all models of $H_{10}$ cluster are obtained from the previous studies by Evangelista *et al*.

S2.1.4.1 H$_{10}$ Ring (1D)

The ring model of H$_{10}$ cluster is an extension of the chain model with one extra H-H bond bringing all H atoms close to each other. The ground state total energies using 1roD geminal wavefunctions for ring model (Fig. S9) show a trend like the energies for the chain model (Fig. 10). The 1roD geminal wavefunctions without the OO step fail to significantly capture the uncorrelated energy, thus tend to closely follow the HF curve. This energy difference with respect to FCI is expected to improve by adding singles-like excitations to proposed 1roD wavefunctions which will take care of orbital relaxation through Thouless rotations, thus reducing the energy values.

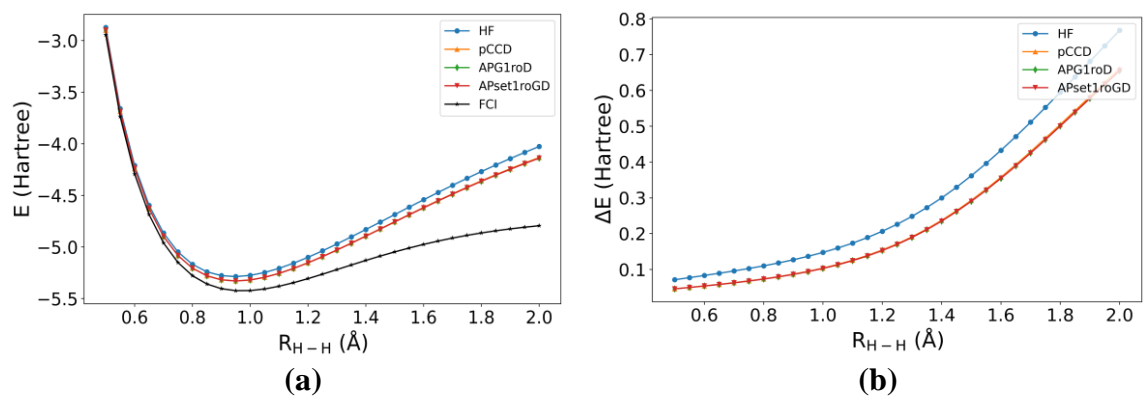

(a)            (b)

**Figure S9.** 1roD: (a) total energies and (b) energy errors ($\Delta E = E_{\text{wavefunction}} - E_{\text{FCI}}$). Ring H$_{10}$ model (STO-6G basis). Also shown, HF and FCI results.

S2.1.4.2 H$_{10}$ Sheet (2D)

Till now we have seen the results of 1roD wavefunctions for H$_{10}$ cluster with simultaneous stretching of 9 and 10 H-H bonds in case of chain and ring, respectively. The sheet model allows us to study the simultaneous breaking of 19 bonds. The overall total ground state energy increases with increasing complexity of the H$_{10}$ models, from chain (Fig. 10), ring (Fig. S9), sheet (Fig. S10) and pyramid (Fig. S9), whereas the overall range for energy error with respect to FCI (range of $\Delta E$ on Y-axis of figures (b)), remains same for each model. This shows the potential of 1roD geminal wavefunctions to robustly study different systems. Just like in case of other H$_{10}$ models, 1roD geminal wavefunctions follow HF curve energy error in the order of 0.1 $E_{\text{h}}$.

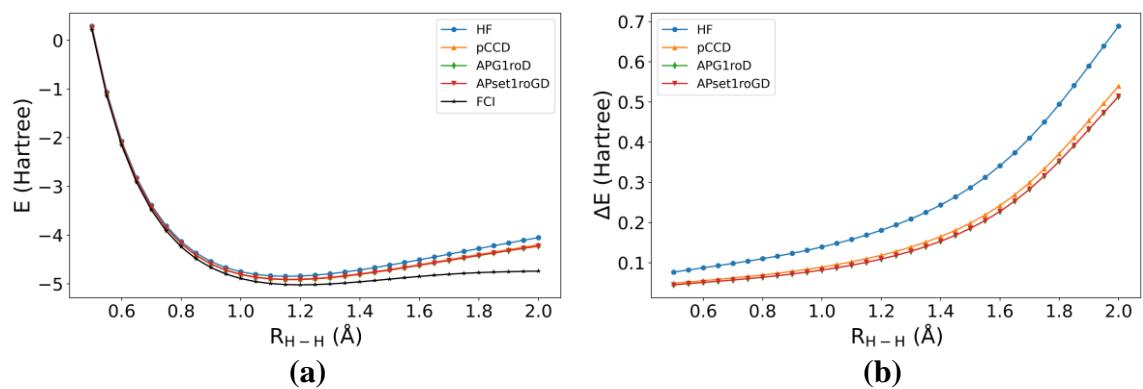

**Figure S10.** 1roD: (a) total energies and (b) energy errors ($\Delta E = E_{\text{wavefunction}} - E_{\text{FCI}}$). Sheet $H_{10}$ model (STO-6G basis). Also shown, HF and FCI results.

S2.2 1-reference orbital geminal wavefunctions with Singles and Doubles: seniority occupied restriction

S2.2.1 Planar $H_4$ models

S2.2.1.1 Trapezoidal model (H4)

The addition of single excitations to 1roD with restrictions in the occupied space shows the significant improvement upon the uncorrelated results. For all H4 models with different bond lengths ($a$), 1roSD_sen-o versions of APG and AP1roG (generalized) give results closer to FCI. The other two wavefunctions, 1roSD_sen-o versions of AP1roG (spin-restricted) and APsetG outperform their counterparts without the S excitations (1roD only) but give higher energy errors compared to the other two wavefunctions.

In case of H4 model with $a = 1.2$ a.u., all 1roSD_sen-o geminal wavefunctions give peak of energy error at $\alpha = 0.05$ (that is $\delta = \pi/20$). For initial geometries while going from square ($\alpha_{\min}$) to linear ($\alpha_{\max}$) conformations, for H4 models with $a = 1.6$ and 2.0 a.u., all 1roSD_sen-o geminal wavefunctions are over optimizing the energy values leading to negative $\Delta E$ (Fig. S11 (d) and (f)). All 1roSD_sen-o geminal wavefunctions give the maximum energy error for H4 models with $a = 1.6$ a.u. and 2.0 a.u. (Fig. S11(d) and (f)) in the order of ~10 m$E_h$. Overall, the inclusion of singles with sen-o condition to 1roD geminal wavefunctions gives an instant improvement upon the energies by 1roD wavefunctions.

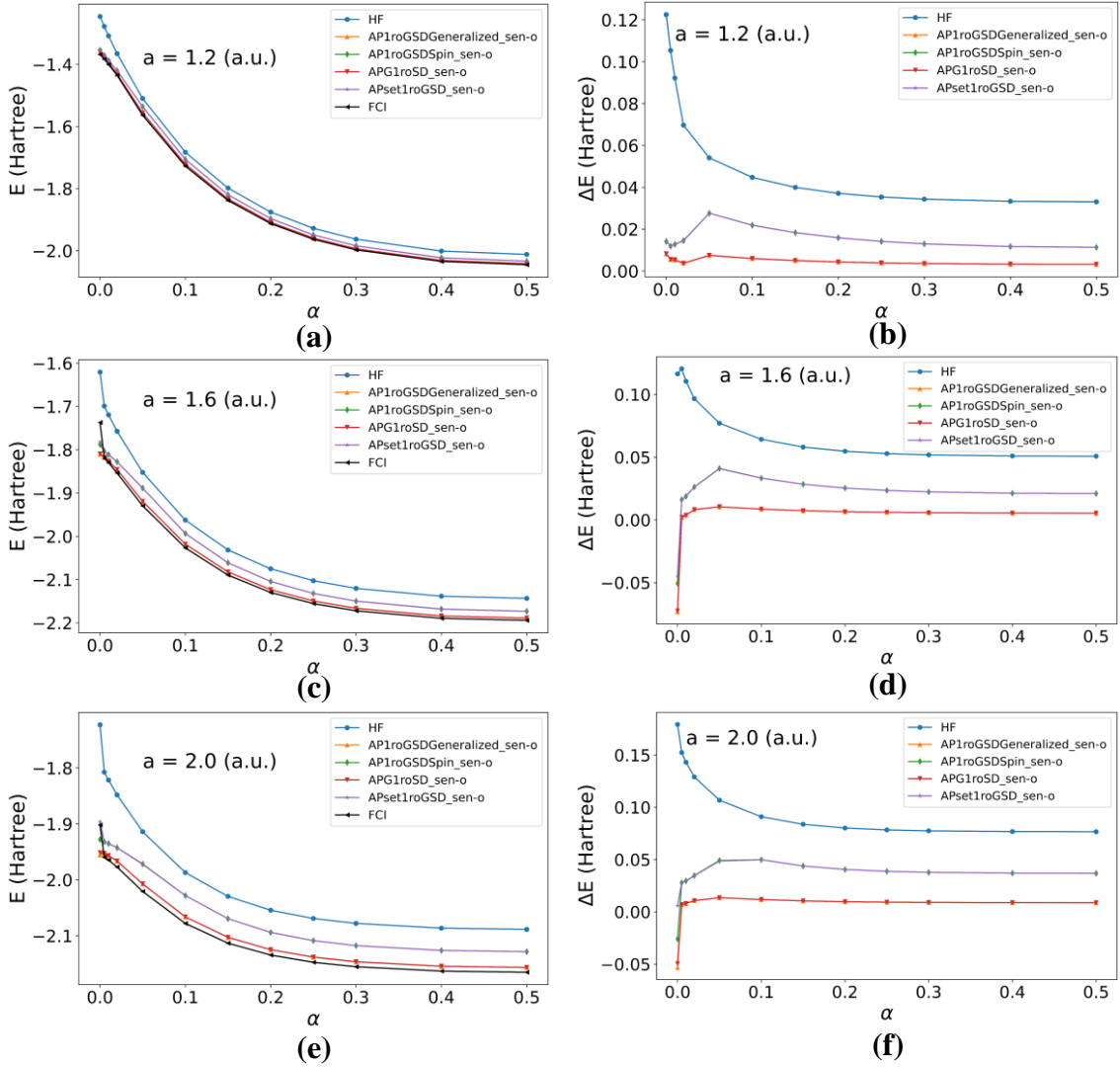

**Figure S11.** 1roSD_sen-o: (a) total energies and (b) energy errors ($\Delta E = E_{\text{wavefunction}} - E_{\text{FCI}}$). Trapezoidal H4 model (STO-6G basis). Also shown, HF and FCI results.

S2.2.1.2 Rectangular model (P4)

For the P4 model, all 1roSD_sen-o wavefunctions give the highest total ground state energy for geometries around $\alpha = a$. From Fig. S12 (b), we see that, for $\alpha = a = 2.0$ a.u., band (i) results give the least energy error with respect to FCI, $\Delta E \sim 6.419$ m$E_h$ compared to all other geometries. For the same geometry, band (ii) result give the energy error of $\Delta E \sim 26.919$ m$E_h$. This difference in $\Delta E$ in band (i) and (ii) highlights the performance of AP1roGSDGenerlalized_sen-o and APG1roSD_sen-o. The discrepancy between the orange and red curves at $\alpha = 2.5$ and $4.0$ a.u. might be coming from the improper energy optimization. For highly stretched H(1)-H(4) and H(2)-H(3), wavefunctions in each band show a maximum overlap with each other.

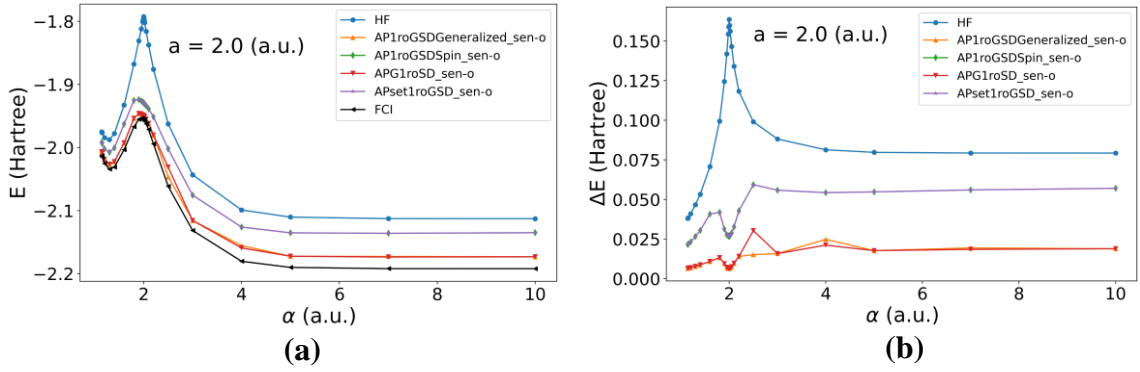

**Figure S12.** 1roSD_sen-o: (a) total energies and (b) energy errors ($\Delta E = E_{\text{wavefunction}} - E_{\text{FCI}}$). Rectangular P4 model (STO-6G basis). Also shown, HF and FCI results.

S2.2.1.3 Linear model (D4)

In linear model D4, we study the stretching of H(2)-H(3) bond by pulling apart two hydrogen molecules by distance H(2)-H(3) = $\alpha$. In Fig. S5 (a), we observed that 1roD geminal wavefunctions give virtually similar energies. After the addition of sen-o to 1roD geminal wavefunctions, 1roSD_sen-o variants of APG (red curve) and AP1roG (generalized) (green curve) wavefunctions presented in Fig. S13 (a) have total energies closer to FCI. For increasing $\alpha$, unlike red and orange curves, green and purple curves maximally deviate away from FCI. The AP1roGSDGeneralized_sen-o and APG1roSD_sen-o wavefunctions give energy values closer to each other, except with a discrepancy at $\alpha$ = 4.0 a.u. and 5.0 a.u.. For $\alpha$ < 2.0 a.u., these two wavefunctions give a minimum error of 2.688 m$E_h$ for $\alpha$ = 0.6 a.u., whereas the energy error increases gradually for $\alpha$ > 2.0 a.u..

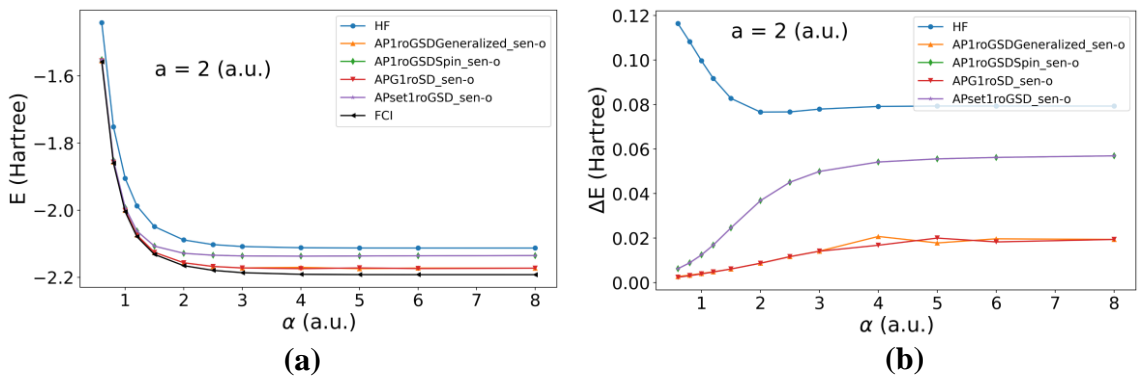

**Figure S13.** 1roSD_sen-o: (a) total energies and (b) energy errors ($\Delta E = E_{\text{wavefunction}} - E_{\text{FCI}}$). Linear D4 model (STO-6G basis). Also shown, HF and FCI results.

S2.2.2 Nonplanar H$_4$ models

S2.2.2.1 T4 model

In the T4 model, we study the geometry of two hydrogen molecules by twisting one H-H bond starting from rectangular conformation leading to highly twisted form with two $H_2$ bonds perpendicular to each other at distance $R$ apart. In Fig. S6, we observed the results for 1roD geminal wavefunctions. Comparing those 1roD results with 1roSD_sen-o (Fig. S14), we observe that the energy error $\Delta E$ decreases for the maximally twisted ($\alpha_{max}$) geometry for T4 conformations with $R < 2.0$ a.u. (Fig. S14(b), (d)). In Fig. S14 (d), for both 1roSD_sen-o variants AP1roG (green and orange cure), the energy error increases for this last geometry. For $\alpha = 0.3$, we see a discrepancy between orange and red curve.

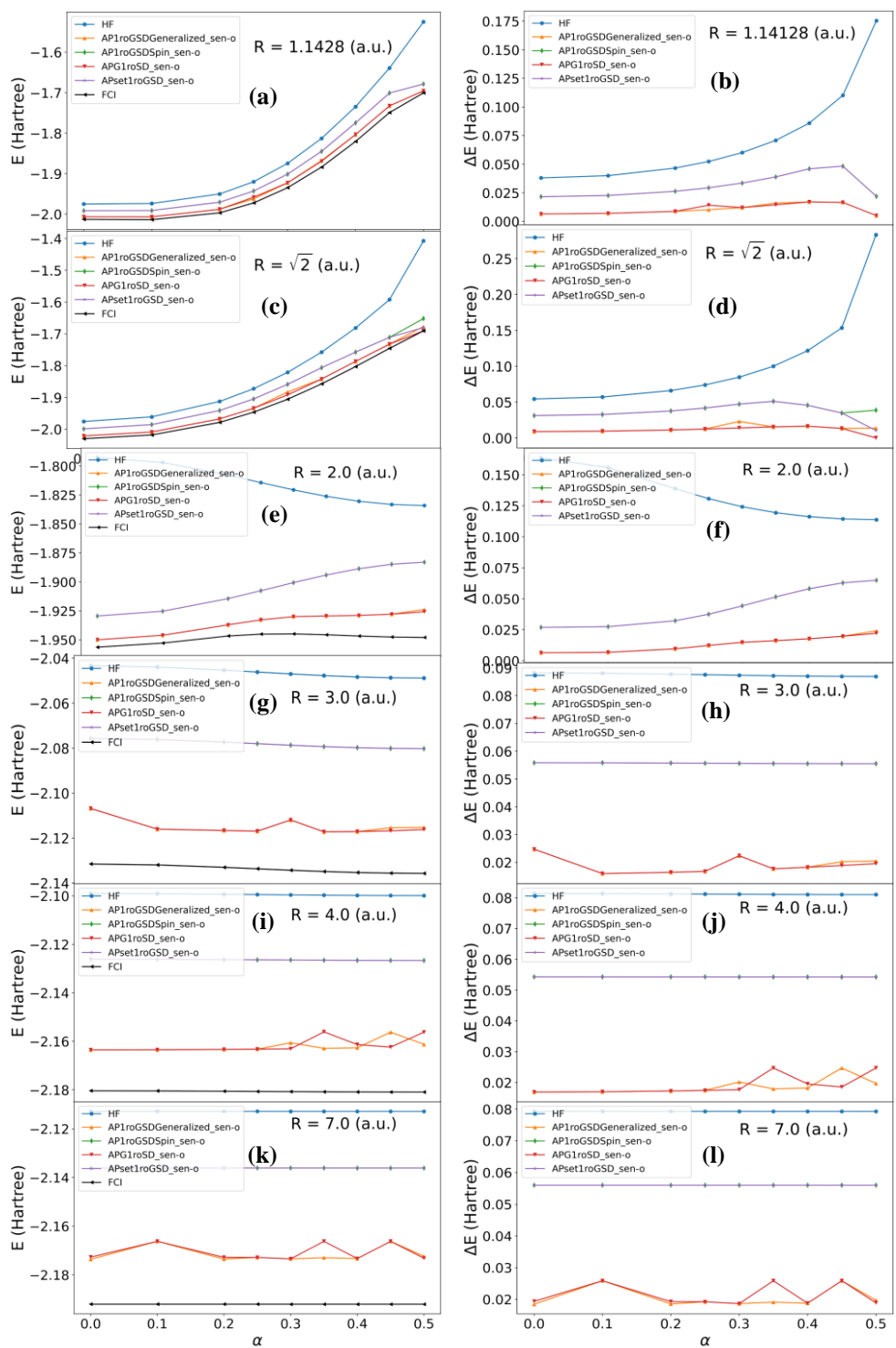

**Figure S14.** 1roSD_sen-o: (a) total energies and (b) energy errors ($\Delta E = E_\text{wavefunction} - E_\text{FCI}$). Nonlinear T4 model (STO-6G basis). Also shown, HF and FCI results.

For $R = a = 2.0$ a.u.: The energy error of 1roSD_sen-o geminal wavefunctions increases with increasing $\alpha$. For $\alpha = 0.5$, all 1roSD_sen-o (Fig. S14 (f)) give the highest error of $\Delta E \sim 24.151$ m$E_h$ for red curve and $\Delta E \sim 64.929$ m$E_h$ for green curve.

For $R = 3.0$ a.u.: The 1roSD_sen-o variants of spin-restricted Ap1roG (green curve) and APsetG (purple curve), Fig. S14 (g), give virtually similar ground state energy around -2.07 $E_h$ as their 1roD counterparts (shown in Fig. S6 (g)). The most accurate methods, 1roSD_sen-o variants of APG (red curve) and AP1roG (generalized) (orange curve) total energy around -2.11 $E_h$ for all geometries, which is closer to FCI.

Similarly, for $R > 3.0$ a.u., we observe that green and purple curves in Fig. S14 (i)-(l) give total energy similar to their 1roD counterparts. But red and orange curves show the impact of addition of singles with sen-o to 1roD wavefunctions showing lower total energies compared to their 1roD counterparts, without performing a separate OO step.

For T4 conformations with $R > 3.0$ a.u., we see some wiggles in energy curves for higher $\alpha$ values, which might be arising from the incomplete optimization of energy values using the BFGS solver.

S2.2.2.2 V4 model

The V4 model allows us to study maximally twisted T4 model, with two H-H bonds of equal length $a = 2.0$ a.u. perpendicular to each other, for different geometries parameterized by the separation between two $H_2$ molecules. In Fig. S7, we observed that 1roD energy values deviate away from the FCI energies with the increasing $\alpha$ value for the V4 model. We notice the similar behavior in the case of green and purple curves in Fig. S15 (a), but red and orange curves give energy values lower than these the former pair of curves giving energies close to FCI.

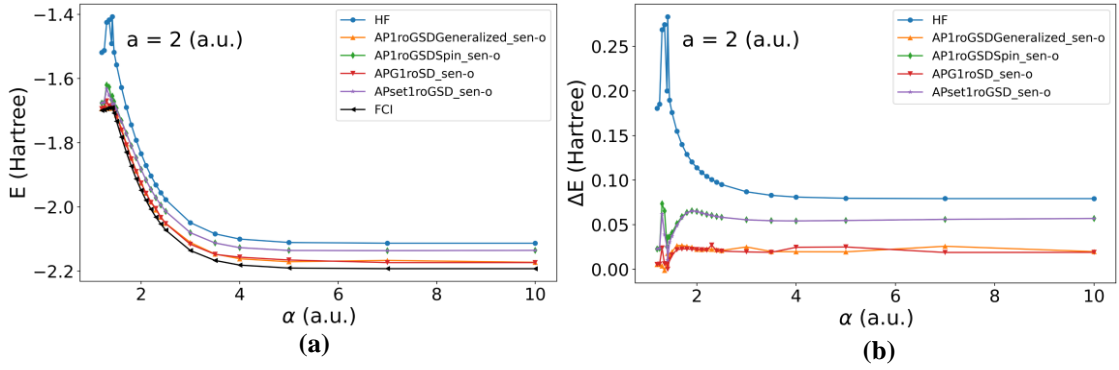

**Figure S15.** 1roSD_sen-o: (a) total energies and (b) energy errors ($\Delta E = E_{\text{wavefunction}} - E_{\text{FCI}}$). Nonlinear V4 model (STO-6G basis). Also shown, HF and FCI results.

The exact solution by FCI gives maximum total energy values of V4 model at $\alpha = \sqrt{2}$ a.u.. We observe that 1roSD_sen-o geminal wavefunctions reach their maximum earlier for different geometries with $\alpha < \sqrt{2}$ a.u.. For $\alpha = 1.35$ a.u., AP1roGSDGeneralized_sen-o overoptimizes the total energy value giving energy lower than FCI causing negative energy error.

For $\alpha > 4.0$ a.u.: 1roD wavefunctions give total energy $E \sim -2.13$ $E_h$. After addition of singles with sen-o to the 1roD wavefunctions, the corresponding variants of AP1roG (generalized) and APG give total energies around $E \sim -2.17$ $E_h$. Whereas 1roSD_sen-o variants of spin restricted AP1roG and APsetG do not show much improvement in their 1roD counterparts.

### S2.2.3 H$_8$ Cube

The 1roD geminal energies deviate away from FCI for higher H-H stretching for the cube model whereas the 1roSD_sen-o geminal wavefunctions (Fig. S16) show an improvement indicating a proper dissociation of all 8 H atoms. The APG1roSD_sen-o wavefunction being most accurate gives energy values with the least energy error. AP1roGSDGeneralized_sen-o results show discrepancies for $R_{\text{H-H}} > 1.2$ Å. For $R_{\text{H-H}} < 1.5$ a.u., the 1roSD variants of spin restricted AP1roG (green curve) and APset1G (purple curve) give energy errors higher than the other two wavefunctions, whereas both variants of AP1roG give higher energy errors for $R_{\text{H-H}} > 1.5$ a.u.. For highly dissociated H-H bonds of the cube ($R_{\text{H-H}} > 3.0$ a.u.,), the orange curve representing the generalized variant of AP1roG converge to the FCI whereas the spin-restricted variant deviate away from the FCI.

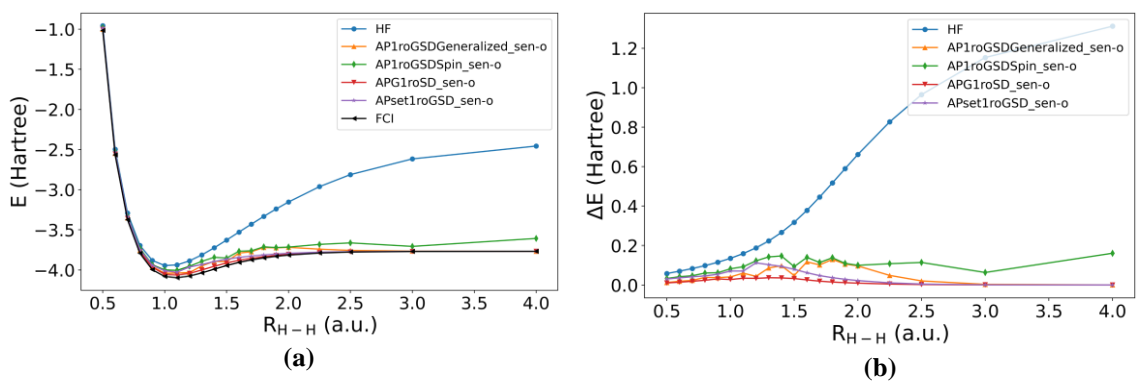

**Figure S16.** 1roSD_sen-o: (a) total energies and (b) energy errors ($\Delta E = E_{\text{wavefunction}} - E_{\text{FCI}}$). Cube $H_8$ model (STO-6G basis). Also shown, HF and FCI results.

S2.2.4 $H_{10}$ Cluster

S2.2.4.1 $H_{10}$ Ring (1D)

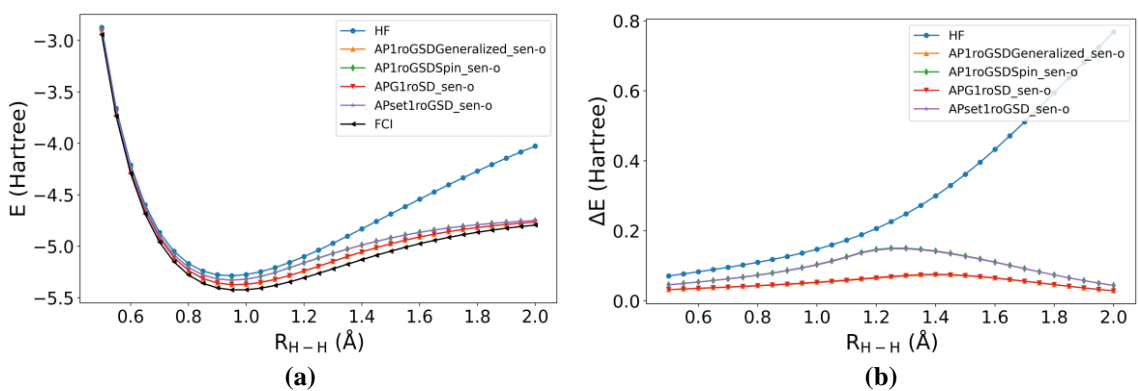

**Figure S17.** 1roSD_sen-o: (a) total energies and (b) energy errors ($\Delta E = E_{\text{wavefunction}} - E_{\text{FCI}}$). Ring $H_{10}$ (STO-6G basis). Also shown, HF and FCI results.

The results from the ring model of $H_{10}$ cluster are like the results from the chain model, except the energy curves for the ring (Fig. S17 (a)) are smoother than the chain model. The most accurate methods, 1roSD_sen-o variants of APG1(red) and AP1roG (generalized) (orange) give maximum energy errors around ring geometries with $R_{\text{H-H}}$ = 1.4 Å. Overall, we observe that the orange and red curves, virtually the same energies. Similarly, the green and purple curves give similar results with total ground state energies higher than the former pair.

S2.2.4.2 $H_{10}$ Sheet (2D)

Going from ring model to sheet model of $H_{10}$ cluster leads to a study of symmetric bond stretching of a 2D structure. From Fig. S18 (b), we can see that red, orange and purple curves show similar trend in energies, giving maximum energy error around $R_{H-H}$ = 1.5 Å, further the energy converge towards FCI. Here, the spin restriction on 1roSD_sen-o version of AP1roG (green curve) causing difficulty in converging to FCI.

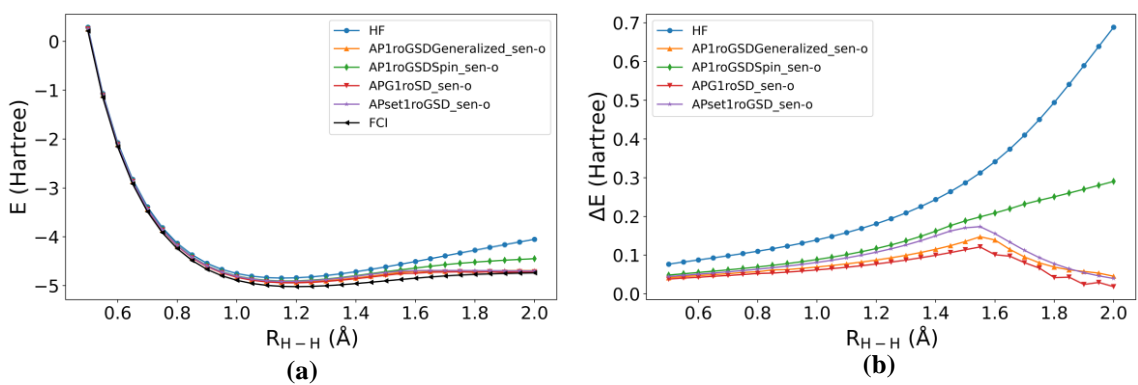

**Figure S18.** 1roSD_sen-o: (a) total energies and (b) energy errors ($\Delta E = E_{\text{wavefunction}} - E_{\text{FCI}}$). Sheet $H_{10}$ (STO-6G basis). Also shown, HF and FCI results.

S2.3 1-reference orbital geminal wavefunction with Singles and Doubles: other seniority patterns

S2.3.1 Planar $H_4$ models

S2.3.1.1 Trapezoidal model ($H_4$)

We observed that the addition of sen-o singles shows a significant improvement upon the results of 1roD geminal wavefunction. The sen-v, sen-ov, and sen-free variants of the 1roSD geminal wavefunctions show trends in energy values similar to the results from the sen-o variants.

In the plots for energy difference with respect to FCI (right column) in Fig. S19, apart from the top curve representing HF energy error, we observe three different bands of overlapping curves representing non-geminal wavefunctions energy error. The lowest set of overlapping curves corresponding to sen-v, sen-ov variants of APG and AP1roG (generalized) 1roSD wavefunctions give the least energy error with respect to FCI, whereas the sen-free variants of these wavefunctions give the higher energy error than the former set of wavefunctions. Because of the spin restrictions, APsetG and AP1roG (spin restricted) 1roSD wavefunctions give highest energy errors for sen-v, sen-ov and sen-free seniority restricted variants.

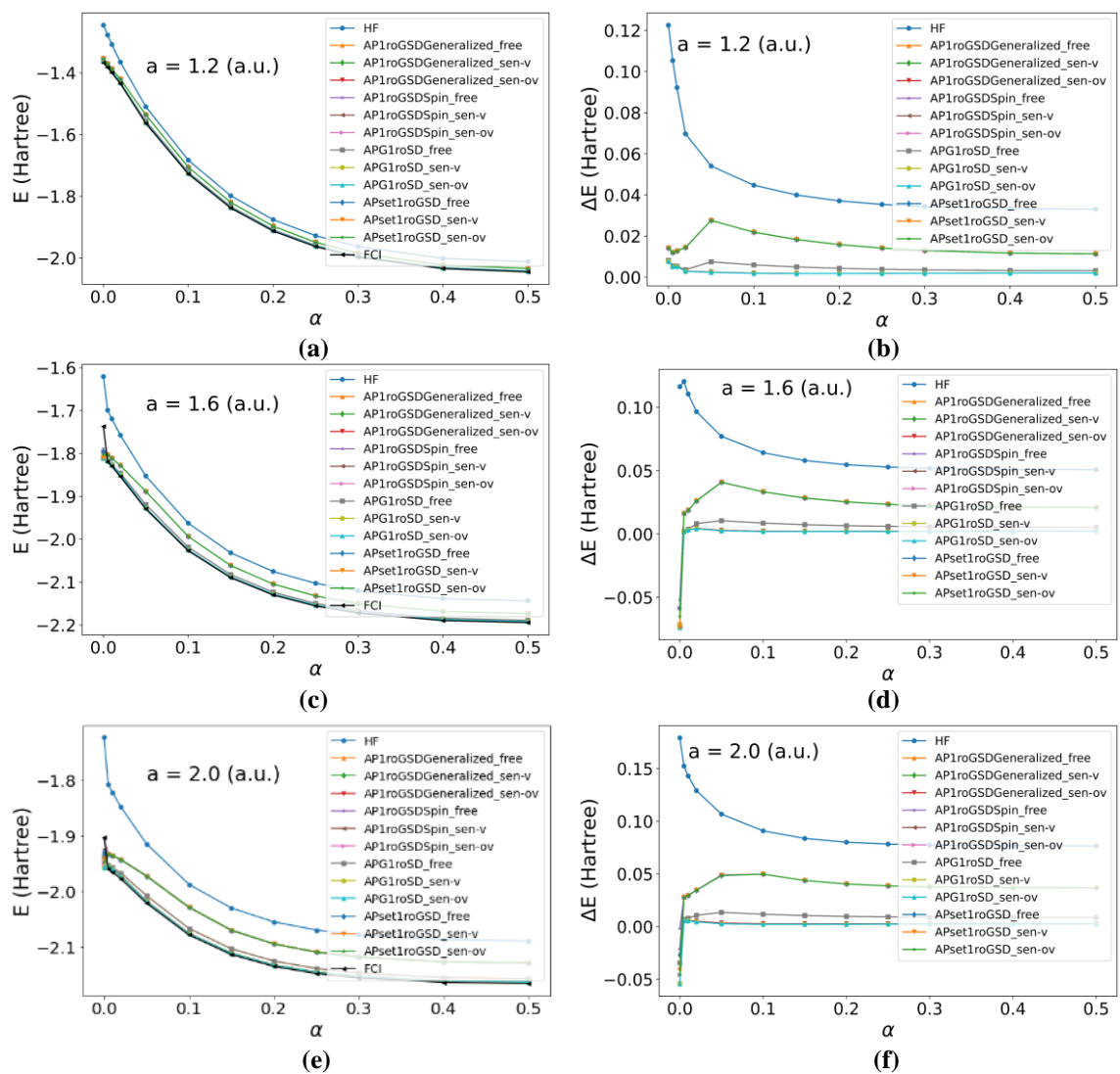

**Figure S19.** 1roSD_sen-ov, sen-v, and sen-free: (a) total energies and (b) energy errors ($\Delta E = E_{\text{wavefunction}} - E_{\text{FCI}}$). Trapezoidal H4 model (STO-6G basis). Also shown, HF and FCI results.

S2.3.1.2 Rectangular model (P4)

In Fig. S20 (b), we observe overlapping results corresponding to the similar set of wavefunctions as for the H4 model (Fig. S19 (b)). Here also the sen-free, sen-v, and sen-ov variants of AP1roG (spin restricted) and APsetG 1roSD give higher total ground state energy values for P4 model. The sen-free variants of the AP1roG (generalized) and APG 1roSD wavefunctions give results close to FCI for $\alpha < 2.5$ a.u., whereas they give energy error of ~ 20 m$E_h$ for the geometries with $\alpha > 2.5$ a.u. The sen-v and sen-ov variants of the AP1roG (generalized) and APG 1roSD wavefunctions give results closest to FCI with maximum $\Delta E$ ~ 5.923 m$E_h$ at $\alpha = 2.05$ a.u..

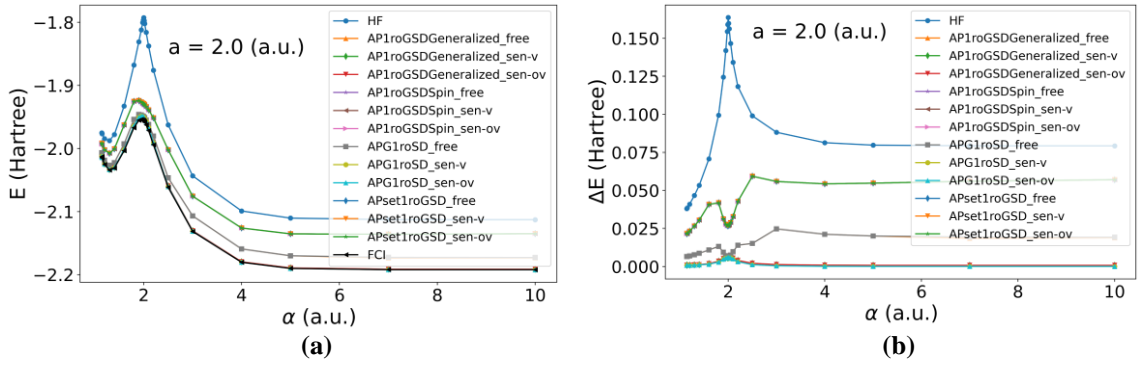

**Figure S20:** 1roSD_sen-ov, sen-v, and sen-free: (a) total energies and (b) energy errors ($\Delta E = E_{\text{wavefunction}} - E_{\text{FCI}}$). Rectangular P4 model (STO-6G basis). Also shown, HF and FCI results.

S2.3.1.3 Linear model (D4)

For the linear D4 model as well, we observe similar relations of virtually overlapping energies of similar set of wavefunctions (Fig. S21). The sen-v and sen-ov variants of the APG and AP1roG (generalized) wavefunctions give maximum error ~ 4.066 m$E_h$ at $\alpha$ = 1.2 a.u., whereas they give virtually constant energy error for $\alpha$ > 2.0 a.u.. The sen-free variants of these two wavefunctions give higher energies than their sen-v and sen-ov counterparts, for $\alpha$ < 3.0 a.u.., they give similar energy values, but we observe some discrepancies in the results for highly stretched D4 model. The rest of the wavefunctions show increasing energy error with increase in $\alpha$.

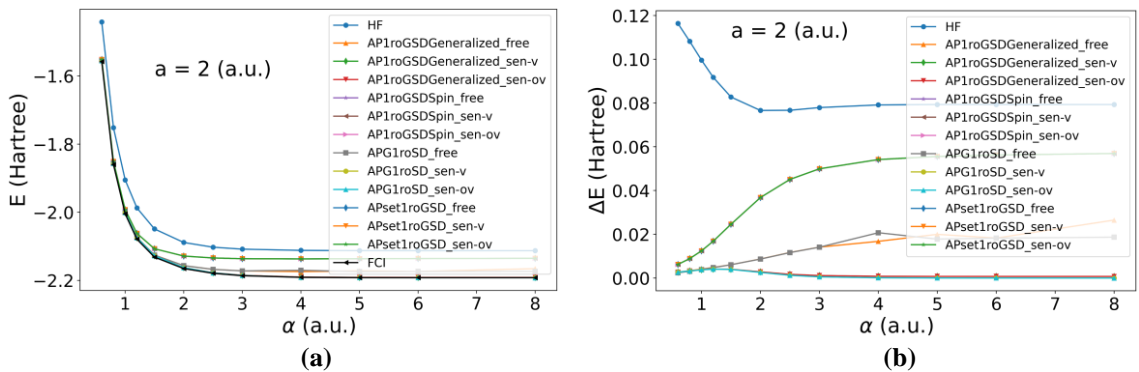

**Figure S21.** 1roSD_sen-ov, sen-v, and sen-free: (a) total energies and (b) energy errors ($\Delta E = E_{\text{wavefunction}} - E_{\text{FCI}}$). Linear D4 model (STO-6G basis). Also shown, HF and FCI results.

S2.3.2 Nonplanar H$_4$ models

S2.3.2.1 T4 model

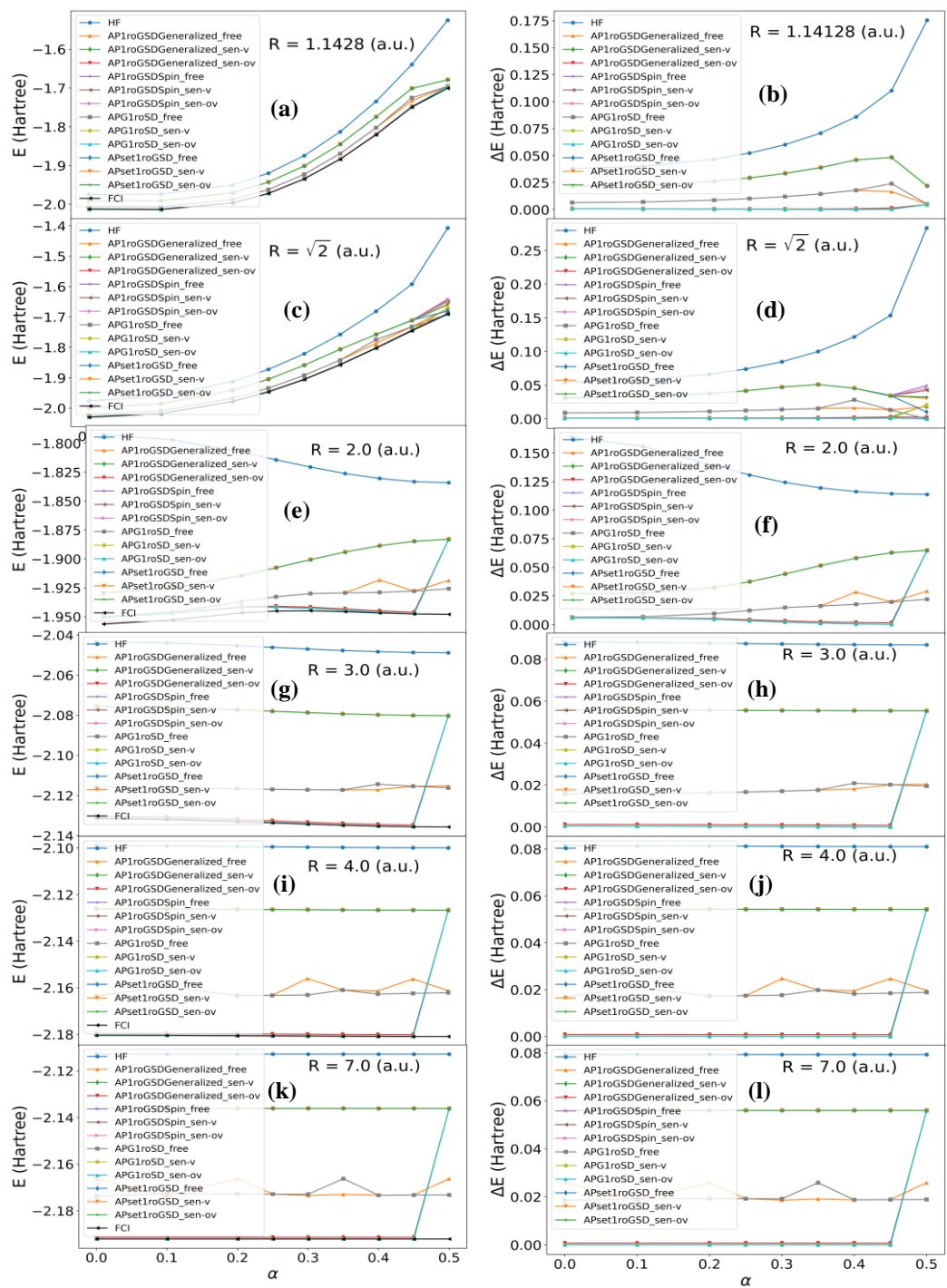

**Figure S22.** 1roSD_sen-ov, sen-v, and sen-free: (a) total energies and (b) energy errors ($\Delta E = E_{\text{wavefunction}} - E_{\text{FCI}}$). Nonlinear T4 model (STO-6G basis). Also shown, HF and FCI results.

In Fig. S22, we see that with increasing internuclear separation $R$ of both the hydrogen molecules (plots from top to bottom), the range of energy error of non-geminal wavefunctions with respect to the FCI is decreasing. For $R \leq a$, sen-v, sen-ov and sen-free variants of the 1roSD wavefunctions show increasing energy error with increasing twist in the conformation (from left to right in each plot) of the T4 model. For $R > a$, we don't see much change in the total ground state energy of the non-geminal wavefunctions with increasing $\alpha$. Just like results from previous model systems, we observed three sets of overlapping results of these wavefunctions. For $R > a$, most of these wavefunctions seem converge toward FCI for the highly twisted T4 model ($\alpha = 0.5$). For the same geometry, all of the wavefunctions here have difficulty in convergence, causing the energy values to deviate away from the FCI, especially in the case of sen-v and sen-ov variants of APG and AP1roG (generalized).

S2.3.2.2 V4 model

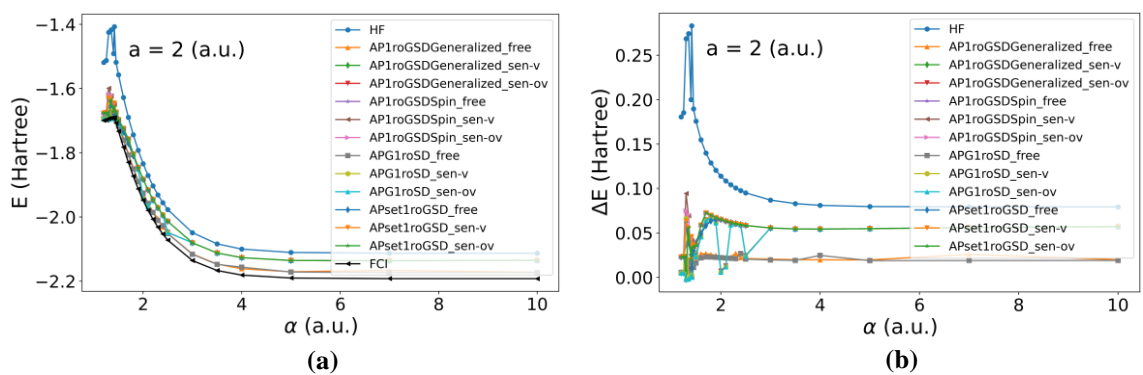

**Figure S23.** 1roSD_sen-ov, sen-v, and sen-free: (a) total energies and (b) energy errors ($\Delta E = E_{\text{wavefunction}} - E_{\text{FCI}}$). Nonlinear V4 model (STO-6G basis). Also shown, HF and FCI results.

Fig. S23 shows the total ground state energy and energy error with respect to FCI for V4 model, the highly twisted T4 model, using the non-geminal wavefunctions. Unlike results from previous models, in case of V4, the sen-free variants of the AP1roG (generalized) and APG give results close to FCI compared to other non-geminal wavefunctions. For V4 conformations, the exact total ground-state energy gradually increases and is highest for $\alpha = \sqrt{2}$ a.u.. This trend in energies can be observed in results by the non-geminal wavefunctions but each wavefunction gives energy peak at different $\alpha$ values leading to multiple wiggles for the initial conformations of V4 model.

S2.3.3 H$_8$ Cube

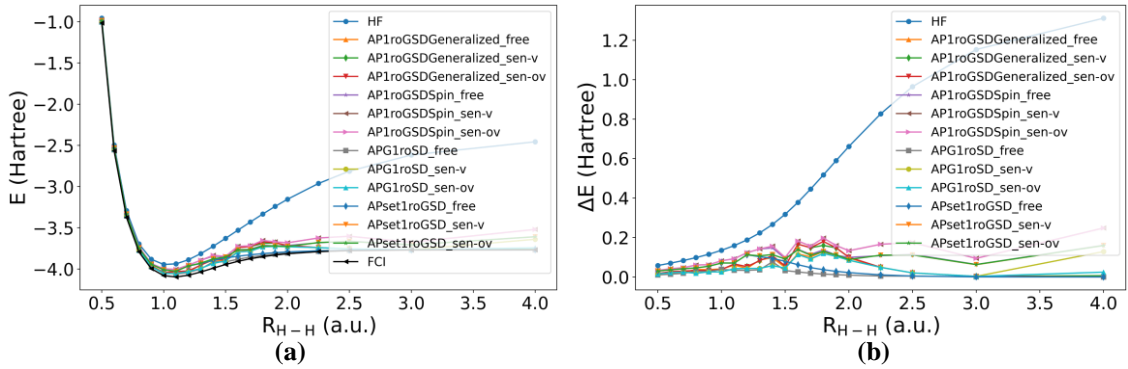

**Figure S24.** 1roSD_sen-ov, sen-v, and sen-free: (a) total energies and (b) energy errors ($\Delta E = E_{\text{wavefunction}} - E_{\text{FCI}}$). Cube $H_8$ model (STO-6G basis). Also shown, HF and FCI results.

In Fig. S24 (b), the gray curve representing the APG1roSD_free wavefunction gives the least energy error compared to other wavefunctions for all geometries of $H_8$ cube, unlike the results from previous model systems. Unlike previous model systems, we observed that sen-v and sen-ov variants of APG1roSD give results tend to deviate away from the exact solution for geometries $R_{\text{H-H}} > 1.5$ a.u..

S2.3.4 $H_{10}$ Cluster

S2.3.4.1 $H_{10}$ Ring (1D)

In the case of $H_{10}$ ring, we see a similar trend in energy values resulting from the non-geminal wavefunctions as from the results for $H_{10}$ chain model. The main difference between Fig. 10 and S25 (b) is that the results for the ring model do not show any discrepancies unlike the chain model. This shows that the nongeminal wavefunctions are easy to optimize for the ring model compared to the chain model using BFGS solver.

S2.3.4.2 $H_{10}$ Sheet (2D)

The sheet model of the $H_{10}$ cluster has more interactions than the ring model. The increasing complexity of geometry leads the non-geminal wavefunctions to be insufficient to converge to the exact solution leading discrepancies in the results for highly stretched geometries. It is interesting to see that the sen-free variants of AP1roG (generalized), APG and APsetG 1roSD wavefunctions show gradual decrease in energy difference with respect to FCI for $R_{\text{H-H}} > 1.55$ Å, the rest of the wavefunctions show increasing energy difference for highly stretched geometries.

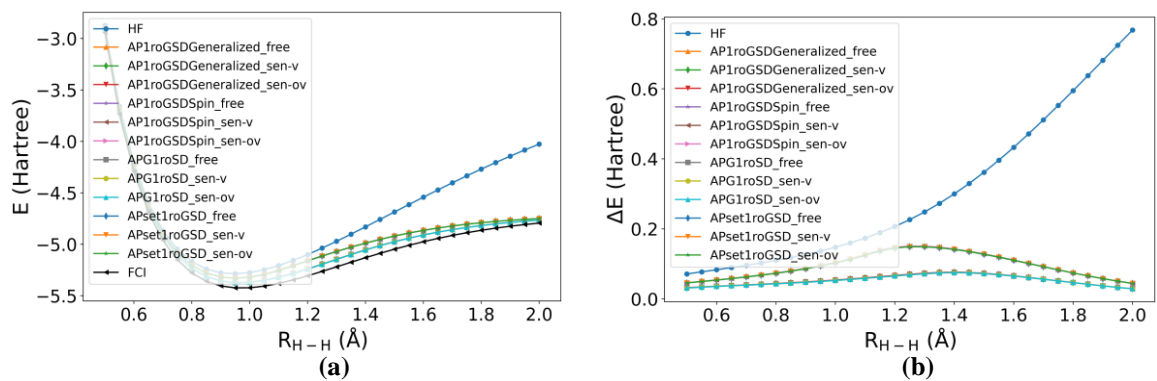

**Figure S25:** 1roSD_sen-ov, sen-v, and sen-free: (a) total energies and (b) energy errors ($\Delta E = E_{\text{wavefunction}} - E_{\text{FCI}}$). Ring $H_{10}$ (STO-6G basis). Also shown, HF and FCI results.

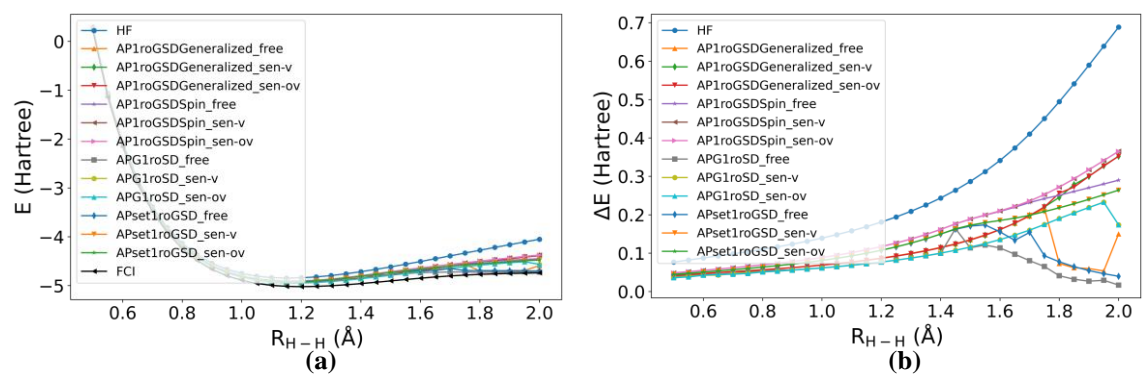

**Figure S26.** 1roSD_sen-ov, sen-v, and sen-free: (a) total energies and (b) energy errors ($\Delta E = E_{\text{wavefunction}} - E_{\text{FCI}}$). Sheet $H_{10}$ (STO-6G basis). Also shown, HF and FCI results.